\documentclass[12pt]{article}
\usepackage[a4paper,%bindingoffset=0.2in,
left=2cm,right=2cm,top=2cm,bottom=2cm,%
footskip=0.25in]{geometry}
\usepackage[english]{babel}
\usepackage{hyperref}
\usepackage{pdfpages}
\usepackage[tc]{titlepic}
\usepackage{xcolor}
\usepackage{graphicx}
\usepackage{url}
\hypersetup{pdfborder=0 0 0}
\usepackage{authblk}
\usepackage{subfigure}
\usepackage{natbib}
\usepackage{amsmath}
 
\title{Modelling the Spatially Varying Non-Linear Effects of Heat Exposure}

\author[1]{Xinyi Chen}
\author[1]{Marta Blangiardo}
\author[1]{Connor Gascoigne}
\author[2,$\ast$]{Garyfallos Konstantinoudis}
\affil[1]{MRC Centre for Environment and Health, School of Public Health, Imperial College London, London, UK}
\affil[2]{Grantham Institute for Climate Change and the Environment, Imperial College London, London, UK}

\affil[$\ast$]{ Corresponding author. Garyfallos Konstantinoudis, Grantham Institute for Climate Change and the Environment, Imperial College London, London, UK. \href{Email:g.konstantinoudis@imperial.ac.uk}{g.konstantinoudis@imperial.ac.uk}}

\date{}  % Remove date

\begin{document}
\maketitle
\newpage
\abstract{
Exposure to high ambient temperatures is a significant driver of preventable mortality, with non-linear health effects and elevated risks in specific regions. 
To capture this complexity and account for spatial dependencies across small areas, we propose a Bayesian framework that integrates non-linear functions with the Besag, York, and Mollie (BYM2) model. 
Applying this framework to all-cause mortality data in Switzerland, we quantified spatial inequalities in heat-related mortality. 
We retrieved daily all-cause mortality at small areas (2,145 municipalities) for people older than 65 years from the Swiss Federal Office of Public Health and daily mean temperature at 1km$\times$1km grid from the Swiss Federal Office of Meteorology. 
By fully propagating uncertainties, we derived key epidemiological metrics, including heat-related excess mortality and minimum mortality temperature (MMT). 
Heat-related excess mortality rates were higher in northern Switzerland, while lower MMTs were observed in mountainous regions. 
Further, we explored the role of the proportion of individuals older than 85 years, green space, average temperature, deprivation, urbanicity, {\color{black} air pollution,} and language regions in explaining these discrepancies. 
We found that spatial disparities in heat-related excess mortality were primarily driven by population age distribution, green space, and vulnerabilities associated with elevated temperature exposure.
}


% \boxedtext{
% \begin{itemize}
% \item Key boxed text here.
% \item Key boxed text here.
% \item Key boxed text here.
% \end{itemize}}

\maketitle
\newpage
\section{Introduction}
Exposure to high ambient temperatures can result in increased illness and mortality \citep{EBI2021698,2022Switz}. The relationship between heat exposure and health is non-linear and typically follows a J-shaped curve \citep{multicountryAmbient, tsDesign, ts_ttlGas, MARTINEZSOLANAS2021e446}, indicating that whilst mild-to-moderate heat may have limited impact, extreme heat can lead to a sharp rise in adverse health effects. The short-term effect of heat exposure on mortality is commonly assessed using a time series approach which allows for non-linearity in the relationship, commonly captured using splines  \citep{multicountryAmbient,tsDesign,ts_ttlGas,MARTINEZSOLANAS2021e446}, and accounts for effect modifiers through stratification or meta-analyses \citep{HeatMort2022Eu,2022Switz,multicountryAmbient,Masselot854,DLNMgaspA,multivarMeta_nonlinear,tsDesign}. 

Previous studies have reported individual and spatial vulnerabilities due to the effect of heat. 
Studies have shown how older populations and individuals with chronic conditions, such as cardiorespiratory diseases and diabetes, are more vulnerable to heat exposure due to lower efficiency in body temperature adaptation \citep{systematicReview_Vulner,HeatMort2022Eu}. 
The relationship has also been found to vary by sex, which can be explained by variation in physical, socio-cultural or lifestyle factors \citep{HeatMort2022Eu}. Studies focusing on quantifying the spatial vulnerabilities of heat exposure have reported that the effect of heat can be modified by local characteristics, such as deprivation, access to medical healthcare services, green space and urbanicity \citep{systematicReview_Vulner,GASPARRINI2022e557}. 
{\color{black}
Most of these studies account for the spatial correlation of temperature effects at a small-area level, assuming that it is governed by certain meta-predictors \citep{GASPARRINI2022e557}. 
However, this cannot capture unmeasured or unmeasurable spatially structured confounding, which is likely resulting in imprecise point estimates. }

{\color{black}
Whilst some studies have accounted for the spatial dependence of the temperature effect using Bayesian spatio-temporal models, most have considered a linear threshold model, which cannot fully accommodate a non-linear relationship \citep{COPD_uk, bennet2014}. 
More recently, non-linear spatially varying effects have been modelled at the neighbourhood level in Barcelona within a Spatial Bayesian distributed lag non-linear (SB-DLNM) approach \citep{sb-dlnm}, without accounting for residual spatial variation or population size, factors that may confound the estimated relationship, nor examining the roles of spatial effect modifiers. 
In addition, the commonly used conditional autoregressive priors remain unscaled \citep{Leroux2000EstimationOD}, complicating interpretation and limiting generalisability to other contexts \citep{BYMvsLeroux}. 
Scaling the spatial prior is important to facilitate hyperprior comparison across different studies. 
Such comparison is particularly important in the context of heat and health as there is increasing literature with data available in multiple cities and countries \citep{HeatMort2022Eu,Masselot854}.
These limitations motivate the need for scalable models that can accommodate flexible spatial structures and produce interpretable, policy-relevant quantities across multiple regions.}

In this project, we propose a flexible model within a Bayesian framework, which is able to account for the spatial variation of the temperature effects, as well as for the spatio-temporal residual confounding. 
{\color{black} 
We extend and develop the results in three ways: 
The most substantial advancement over previous work is the expansion to a nationwide analysis, covering 2,145 municipalities across all 26 cantons of Switzerland from 2011 to 2022. This larger spatial scale enables a detailed investigation of geographic variation in heat-related mortality and the role of effect modifiers.}
{\color{black}
We implement our model using the Integrated Nested Laplace Approximation (INLA) algorithm \citep{inlaref}, which offers substantial advantages over traditional Markov Chain Monte Carlo (MCMC) approaches. 
INLA is significantly more computationally efficient, allowing us to scale our models to large datasets with feasible run times. 
Moreover, INLA is a user-friendly framework with built-in functionality well-suited to our analytical needs. 
For example, it supports structured priors - such as Gaussian priors on spatial fields combined with penalised complexity (PC) hyperpriors - that promote model identifiability and penalise overfitting, making it particularly suitable for hierarchical and spatially structured models. 
Finally, we derive a set of epidemiological metrics relevant for results dissemination which we then use in a secondary analysis to quantify the effects of observed and unobserved environmental factors in the estimated spatial variations in the effects of heat.  }

The remainder of the paper is structured as follows:  
In Section \ref{Data}, we describe the dataset used for the analyses. 
In Section \ref{method}, we introduce the statistical approach to account for the spatial variation and spatial dependence of the non-linear heat effects across the municipalities, the calculation methods of epidemiological metrics, and the model to conduct the secondary spatial effect modifier analysis. 
In Section \ref{result}, we introduce different complementary epidemiological metrics derived from our model framework and the results of the secondary spatial effect modifier analysis, which can be used to obtain a comprehensive picture of the effects of heat on all-cause mortality in Switzerland.
Finally we discuss the results and make concluding remarks in Section \ref{discussion}. 
{\color{black}
We provide the complete code in \url{https://github.com/fxinyichen/SwissHeat_svc} and the simulated data set in \url{https://doi.org/10.5281/zenodo.16923676}.}

\section{Data description}\label{Data}
{\color{black} 
To facilitate comparison with previous studies in Switzerland, we are focusing on the summer months from June to August \citep{2022Switz}.} 
We retrieved the daily all-cause mortality data and population data between 1st of June and 31st of August each year covering 2011-2022 in 2,145 municipalities within 26 cantons in Switzerland from the Swiss Federal Office of Public Health \citep{SwissPop}.
A map of Switzerland and its cantons is shown in Supplementary Figure 1. We focused on the summer months due to our interest in heat-related exposures, but the algorithm can be extended to capture the cold-related effects and incorporate longer lags \citep{DLNMgaspA,multicountryAmbient}.
As only yearly population is available (on the 31st of December for each year), we used linear interpolation by municipality to estimate the daily summer population from 2011 to 2022. 
Based on the results of \citet{HeatMort2022Eu}, we restricted our analysis to individuals aged over 65 years, as they experience higher vulnerability to heat exposure.

During the same time period, we retrieved the daily mean temperature at 1km$\times$1km grid from the Federal Office for Meteorology and Climatology \citep{SwissTempData}. We obtained average daily municipality temperature by taking a population-weighted  average of the grids that cover each municipality, as described previously \citep{tempGrid}.
We accounted for up to a 3-day lagged exposure by taking the mean of the temperature exposures across lags 0-3 \citep{heatLag03,DLNMgaspA,bennet2014}. We adjusted the analysis for national holidays \citep{NagerDate}, day of week, year (to represent long-term trends) and day of year (to represent seasonality).

We considered the following spatial effect modifiers: language region \citep{Swiss_lang}, urbanicity \citep{Swiss_urban}, deprivation \citep{Swiss_socioeconomic}, green space \citep{Swiss_NDVI}, average temperature across the period, {\color{black}air pollution \citep{Swiss_NO2}}, and proportion of people older than 85 years old. 
We classified the language region into three groups: German, French, and Italian, reflecting different lifestyle factors across Switzerland. 
For urbanicity, we defined three categories: rural areas (representing thinly-populated rural regions), semi-urban areas (consisting of towns and suburbs or small urban area), and urban areas (including cities and large urban regions). 
This captures different temperature exposures across the different urban settings. 
To account for deprivation, we categorized the continuous socioeconomic position (SEP) data, which is a Swiss-based index strongly associated with household income \citep{Swiss_socioeconomic}, into three groups: low SEP (1st quartile), baseline SEP (2nd and 3rd quartile), and high SEP (4th quartile).
As a measure of green space, we used the Normalized Difference Vegetation Index (NDVI) available for 2011 and 2013-2018  \citep{Swiss_NDVI}. We averaged the available NDVI data to represent the green space coverage from 2011 to 2022. 
To account for potential adaptation mechanisms to higher temperatures, we calculated the mean temperatures across municipalities during the study period. 
{\color{black} 
To capture the effect of air pollution, we selected NO2 as we have it on a spatial resolution compatible with the outcome, and it has been linked to causes of deaths such as cardiovascular and respiratory according to literature \citep{NO2_copd,Meng2021NO2_MortalityCardRespir,Yang2025NOxMortality}.
We used the NO2 concentration (unit: $\mu g/m^3 \times 10$) data in 2010 to represent the NO2 exposure from 2011 to 2022.}
Finally, using the population data, we defined the proportion of population older than 85 years old, as they are expected to be more vulnerable to heat exposure.

\section{Methods}\label{method}
Our approach consists of three steps: (i) we model the effects of temperature on summer mortality in Switzerland; 
(ii) we estimate a range of epidemiological metrics to fully explore the heat-mortality relationship; 
(iii) we disentangle {\color{black}the spatial differences between one of the epidemiological metrics, heat-related excess mortality rate}, and the spatial effect modifiers (language region, urbanicity, deprivation, green space, average temperature, {\color{black}air pollution,} and proportion of people over the age of 85). 
The statistical analyses are performed in R 4.4.1 \citep{R}.

\subsection{Statistical approach}
Let $Y_{dtm}$ be the number of deaths for all causes, for the $d$-th day in $t$-th year, and $m$-th municipality. 
As we restrict our analysis to the summer months (June, July, August), we use $d = 1, \dots, 92$ for the day index, $t = 1, \dots, 12$ for the year (2011 to 2022), and $m = 1, \dots, 2145$ for the municipalities. 

In the first step, we model the relationship between all-cause mortality and temperature using a Poisson likelihood with a log-link function,
\begin{equation*}
\label{main_model}
    \begin{split}
     Y_{dtm}  & \sim  \text{Poisson} (E[Y_{dtm}])\\
     \log(E[Y_{dtm}]) & =  \log(P_{dtm}) + \beta_0 + \mathbf{X}_{dtm}\cdot\mathbf{\beta}_m+  \mathbf{Z}_{dtm} \cdot\mathbf{\gamma}+ \omega_d + \delta_t + b_m, 
    \end{split}
\end{equation*}
where $P_{dtm}$ is the population for the d-th day, t-th year, and m-th municipality; 
$\beta_0$ is the global intercept, representing the baseline log mortality rate; 
$\mathbf{X}_{dtm} \cdot \mathbf{\beta}_m$ is a non-linear function of the average daily temperature across lags 0-3 represented by a matrix of basis functions, $\mathbf{X}_{dtm}$, and its coefficients, $\mathbf{\beta}_m$, which are the combination of fixed effects $\mathbf{\beta}$ and random effects $\mathbf{\beta\prime}_m$ of the temperature on mortality, (i.e., $\mathbf{\beta}_m  = \mathbf{\beta}  + \mathbf{\beta\prime}_m$);
$\mathbf{Z}_{dtm}$ is a matrix representation of a set of categorical spatiotemporal confounders (the day of week and whether or not it is a public holiday), with coefficients $\mathbf{\gamma}  = [\gamma_1, \gamma_2, \cdots, \gamma_7]^T$, where $\gamma_i$ ($i =1,2,\cdots,6$) are coefficients for the categorical variable \textit{day of week} with Sunday as the reference level and $\gamma_7$ is the coefficient for the binary variable \textit{holiday} with non-holiday as the reference level. 
The final three terms $\omega_d$, $\delta_t$, and $b_m$ are random effects accounting for unmeasured confounding at the level of day (i.e., to model seasonality), year (i.e., to model long-term trends), and space (i.e. to model dependency across municipalities), respectively. 

To account for the non-linear relationship between temperature and mortality, we use the natural cubic splines, commonly employed in the literature for studying the temperature-mortality association \citep{multicountryAmbient,GASPARRINI2022e557,shouldAdjust}. 
Firstly, we select three knots at the 10-, 75- and, 90-th quantiles of the temperature range $D$ and obtain the initial basis matrix $\mathbf{X{\prime}}_{dtm}$ without intercept (this means the value of each basis at the minimum temperature equals zero). 
{\color{black} 
Then, according to an exploratory analysis, we select 12$^\circ C$ as a country-level reference temperature and rescaled the basis matrix to be centered at this value. }
To ensure that the log-relative risk is equal to 0 at the reference temperature, 
we rescale the initial basis matrix to be centered at 12$^\circ C$ by subtracting the values of each of the four basis functions in matrix $\mathbf{X{\prime}}_{dtm}$ at the 12$^\circ C$: 

$${\color{black}\mathbf{X}_{dtm} = \mathbf{X{\prime}}_{dtm} -  \mathbf{1} \cdot\mathbf{x{\prime}}_{[x = 12]} } , $$
{\color{black} 
where $\mathbf{x{\prime}}_{[x = 12]}$
is a four-element vector containing the values of the basis matrix $\mathbf{X{\prime}}_{dtm}$ at 12$^\circ C$.
We can also run the model with the initial basis $\mathbf{X{\prime}}_{dtm}$, and reparameterize the values of logRR, which will be explained in Section \ref{RRsec}, similar to the practice of centering in the frequentist distributed lag non-linear model (DLNM) approach \citep{DLNMgaspA,DLNM_R}.}

\subsubsection{Prior specification}
As we implement the model in a Bayesian paradigm, we specify priors for all parameters. 
We set a $\beta_0 \sim N(0,\infty)$ prior for the intercept and $\mathbf{\beta}$, $\mathbf{\gamma} \sim N(0,1000)$ prior for the fixed effect parameters. 
{\color{black}
As we wish for the coefficients of the basis function to vary in space, we consider a Besag, York, and Mollie (BYM2) model with scaled spatial prior to facilitate the interpretation and generalisiability, as mentioned previously. 
The BYM2 model is a transformed combination of a zero-mean Gaussian prior and an intrinsic conditional autoregressive (ICAR) prior \citep{BYMref}. }
Based on our specification of the basis function, let $\mathbf{\beta\prime} = [\mathbf{\beta\prime_{1}}, \mathbf{\beta\prime_{2}}, \mathbf{\beta\prime_{3}}, \mathbf{\beta\prime_{4}]}$, specifically, and:
\begin{equation*}
    \label{BYM2}
    \mathbf{\beta\prime_{j}} = \sigma_{\beta_j} \left(\sqrt{1-\phi_{\beta_j}} \mathbf{v}_{j\star} + \sqrt{\phi_{\beta_j}} \mathbf{u}_{j\star}\right), \quad j=1,2,3, 4
\end{equation*}
where $\mathbf{v}_{j\star}$ {\color{black} (unstructured)} and $\mathbf{u}_{j\star}$ {\color{black} (spatially structured)} are standardised versions of two Gaussian priors $\mathbf{v}_j$ and $\mathbf{u}_j$ (ICAR) to have variance equal to 1 \citep{BYMvsLeroux}. 
The adjacent matrix for $\mathbf{u}_{j\star}$ was defined using the queen criterion of contiguity, hence areas are neighbours if they share a common edge or a common vertex.
The hyperparameter $\phi_{\beta_j}$, known as the spatial mixing parameter, measures the proportion of marginal variance that can be explained by the structured spatial effect, $\mathbf{u}_{\star}$, \citep{BYMref}. The hyperparameter $\sigma_{\beta_j}$ is the standard deviation of the spatial field. 
For the residual confounding in space ($b_m$), we use a BYM2 prior with the hyperparameters $\sigma_{b}$ and $\phi_{b}$. 

To account for seasonality we define a random walk 2 (RW2) process: 

$$\omega_d|\omega_{d-1}, \omega_{d-2}, \sigma_{\omega}^2
        \sim N(2\omega_{d-1} - \omega_{d-2}, \sigma_{\omega}^2)$$
where $\sigma_{\omega}^2$ is the variance of the process and $\omega_{d-1}, \omega_{d-2}$ the random effect for the two days before $d$.
For the residual confounding due to the long-term trend, we specify a zero-mean Gaussian process, $\delta_t \stackrel{\text{iid}}{\sim}  N(0,\sigma_{\delta}^2)$, with $\sigma_{\delta}$ as the standard deviation. 

\subsubsection{Hyperpriors}
On all the hyperparameters we specify Penalised Complexity (PC) priors. 
PC priors penalise model complexity according to the distance between the base (simpler) model and an alternative (more complex) model \citep{inlaref,Gómezbook,PCpriorBook,BYMvsLeroux,PriorSimpson}. 
{\color{black}
We specify informative priors to reflect our assumptions about the variability and structure of the model components. For the standard deviation of the random effects across year, region, and spatially-varying splines, we assume that the probability of it exceeding 1 (which corresponds to mortality risks greater than exp(2)) is very small, and set this probability to 0.01. For the mixing parameters governing the fields $\beta^{\prime}$ and $b$, we set the prior probabilities $Pr(\phi_{\beta_j} < 0.5) = 0.5$ and $Pr(\phi_{b} < 0.5) = 0.5$,  reflecting our lack of prior knowledge about whether overdispersion or strong spatial autocorrelation should dominate. These prior choices were shown to lead to adequate posterior inference for all-cause mortality in Switzerland \citep{5EUCOVID2020excess}.
For the random effect across day of year, we assume a strong prior by setting the probability that its standard deviation exceeds 0.01 to 0.01. This assumption is imposed to enforce smoothness in the seasonality term while still allowing the temperature effect to capture seasonal variation. We have tested the robustness of this assumption in a sensitivity analysis.
Finally, there is limited prior information on spatial vulnerability to heat in the Swiss context to inform prior selection for the spatially varying splines. However, a previous study using cantons as spatial units found that the maximum discrepancy from the average heat effect was approximately two-fold, supporting the chosen prior specification \citep{2022Switz}. The mixing parameter prior reflects our limited evidence of the spatial smoothness. 
}

The model is implemented using Integrated Nested Laplace Approximation (INLA) in the R package (R-INLA) to ensure a manageable running time \citep{R,inlaref}.
{\color{black} We simulate 1000 INLA posterior samples for the the coefficients in the main model.
}

\subsection{Relative mortality risk of temperature}\label{RRsec}
Firstly, for computational efficiency, we define a set of values (vector $\mathbf{x}$) on the temperature domain $D$, starting from -6.81 to 29.48$^\circ C$, with 200 steps (sampled from the whole temperature exposure data), $\mathbf{x}= [-6.81,6.28, \dots, 29.48]$, derive the corresponding sampled basis matrix $\mathbf{X}_{s}$ and calculate the log relative risk (RR) of temperature-related mortality in each municipality $m$, after accounting for confounding, as follows:

\begin{equation*}
\label{matrixmulti_INLA}
    {\color{black}
    \begin{split}
         \log \text{RR}^{\star}_m(x) & = \mathbf{X}_{s}\cdot \mathbf{\beta}_m , \quad x \in \mathbf{x} \\
         \mathbf{\beta}_m  & = \mathbf{\beta}  + \mathbf{\beta\prime}_m,
    \end{split}
    }
\end{equation*}
{\color{black} where $\mathbf{\beta}$ is the nationwide effect of the coefficients of the basis function, and $\mathbf{\beta\prime}_m$ are the zero-mean spatial deviations.}

{\color{black} 
As the current $\log \text{RR}^\star_m(x)$ is centered at the nationwide reference 12$^\circ C$ across all the municipalities, we then rescale the log relative risk ($\log \text{RR}^\star_m(x)$) on the region-sample-specific minimum mortality temperature (MMT), consistent with previous studies \citep{Masselot854}. 
To achieve this, we retrieve the 1000 logRR sample curves for each municipality and calculate the municipality specific MMT as the temperature with minimum logRR within the 25-90th temperature range of that area in summer months. 
The selection of these percentiles is to provide more stable estimates, and it is in line with the literature \citep{SwissAFANRisk}. 
To rescale on the MMT: }

\begin{equation*}
{\color{black} 
    \log \mathrm{RR}_m(x) = \log  \mathrm{RR}^\star_m(x) - 1 \cdot \log  \mathrm{RR}^\star_{m _{[x=\text{MMT}]}}, \quad \text{where $\log  \mathrm{RR}^\star_{m_{[x = 12]}} =0$}
    }
\end{equation*}

The relative risk can be shown in coarser spatial resolutions that might be of public health relevance, whilst maintaining the full uncertainty. 
For instance, as cantons in Switzerland are responsible for their populations' healthcare and welfare, reporting results at this level could highlight differences due to different policies, preparedness, and resilience to the effect of heat. 
We define the posterior $\text{{\color{black}log relative risk}}$ of the $c$-th canton ($c=1, \dots, 26$) on the set of sampled temperature values $x \in \mathbf{x}$ as:
\begin{equation*}
\label{canton coef}
\begin{split}
   \textcolor{black}{\log \mathrm{RR}^\star_c(x)} & =  \textcolor{black}{\mathbf{X}_{s} \cdot  \mathbf{\beta}_c}\\
     \beta_{jc} & = \sum_k \beta_{jk} \frac{P_{k}}{P_c} , \quad j = 1,2,3,4, \quad k \in M_c\\
     % w_k & = \frac{P_{k}}{\sum_k P_k}, \quad k \in m_c
\end{split}
\end{equation*}
with $M_c$ denoting the set of municipalities that belong to the $c$-th canton, $P_k$ representing the population in municipality $k$ belonging to canton $c$, and $P_c$ is the population of canton $c$.
That is, $P_c = \sum_k P_k$, where $k \in M_c$.
In this way, we retrieve the population weighted coefficients of the basis function, that can be used to estimate the temperature-mortality curve in every canton. 
{\color{black} 
Similar to the municipality-level relative risk, we rescale the $\log \text{RR}^\star_c$ to the cantonal-specific MMT for each canton.}

\subsection{Epidemiological metrics} \label{metric}
\textbf{Minimum mortality temperature (MMT)}: 
{\color{black}
Based on the definition mentioned previously, we derive the median values of MMT across the posterior samples at municipality level. 
We also calculate the MMT percentile (MMP), which is the percentile of the temperature distribution corresponding to the MMT estimated in each region \citep{MMTglobal,Masselot854}.}

\textbf{Excess mortality rates attributable to heat (ERH)}:  
{\color{black} 
The ERH metric represents the proportion of excess death attributable to the heat exposure among a specific population. 
To calculate $\text{ERH}_m$ in the m-th municipality, we first estimate the fraction of deaths attributable to daily temperatures x, denoted $\text{AF}_m(x)$. 
This is obtained from the relative risk at temperature x, $\text{RR}_m(x)$, as shown below. 
Next, we multiply $\text{AF}_m(x)$ by the number of deaths observed at temperature x, $Y_m(x)$, and sum across all temperatures above the MMT. 
This results in the excess death counts due to heat ($\text{ECH}_m$) in the m-th municipality. Finally, we can retrieve $\text{ERH}_m$ through dividing $\text{ECH}_m$ by the population ($P_m$), or alternatively, $\text{AFH}_m$ by dividing by the total number of cases ($Y_m$) in the m-th municipality.}

\begin{equation*}
\begin{aligned}
\mathrm{AF}_m(x) &= \frac{\mathrm{RR}_m(x)-1}{\mathrm{RR}_m(x)}, 
\quad x \in \boldsymbol{x},\; {\color{black}x>\mathrm{MMT}} \\
{\color{black}\mathrm{ECH}_m} &= {\color{black}\sum_{x>\mathrm{MMT}} \mathrm{AF}_m(x)\, Y_m(x)} \\
{\color{black}\mathrm{ERH}_m} &= {\color{black}\frac{\mathrm{ECH}_m}{P_m}} \\
{\color{black}\mathrm{AFH}_m} &= {\color{black}\frac{\mathrm{ECH}_m}{Y_m}}
\end{aligned}
\end{equation*}

{\color{black}
Using the full posterior distributions, we can retrieve the full uncertainty of the derived epidemiological metrics for any spatial and/or temporal aggregation, higher than that of municipality and day. 
That is, for one region, we have 1000 INLA posterior coefficients on the basis function, a full posterior for each relative risk and epidemiological metric. 
For one posterior sample, we retrieve the logRR curves by municipality and calculate the municipality-sample-specific MMT. 
Then, we rescale the logRR to be centered at municipality-sample-specific MMT and calculated the metrics for each sample. 
}

\subsection{Spatial effect modifiers}
To quantify the effect of the selected effect modifiers on the {\color{black} spatial differences in the ERH}, 
we use the ERH metric derived from Section \ref{metric} as the main outcome in the following model, where $c_m$ is the canton which municipality $m$ belongs to. The model for ERH is written,
\begin{equation*}
\label{effectMod_eq}
\begin{split}
    \text{ERH}_{m} & \sim \text{Normal}(\mu_{m}, \sigma_{\text{ERH}}^2) \\
    \mu_{m} &= \alpha_0 +  \mathbf{H}_m  \cdot \mathbf{\alpha} + \zeta_{c_m} + \xi_m 
\end{split}
\end{equation*}
where 
$\alpha_0$ is the baseline excess mortality rate, with prior $\alpha_0 \sim N(0,\infty)$;
$\mathbf{\alpha}$ is a vector of the coefficients of the spatial effect modifiers associated with $\mathbf{H}_m$, the vector of  {\color{black}all} the effect modifiers at area $m$, and $\alpha \sim N(0,1000)$;
$\zeta_{c_m}$ captures the spatial residuals across cantons, with $\zeta_{c_m}\stackrel{\text{iid}}{\sim}  N(0,\sigma_\zeta^2)$,  $P(\sigma_{\zeta}>1) = 0.01$;
and $\xi_m$ captures the spatial residuals across municipalities using a BYM2 prior as defined in the main model.

We show median and 95\% credible intervals (CrIs) of $\mathbf{\alpha}$ using two approaches. 
In the first approach, we do not propagate the uncertainty of ERH, and use its median value for the regression. 
For the second approach, we take 200 samples of the posterior of ERH derived from Section \ref{metric}, fit the above model 200 times, and summarise the posterior distributions of $\mathbf{\alpha}$ across the 200 samples.

\subsection{Sensitivity analysis}
{\color{black} 
To test the robustness of the model, we compare our main results from four aspects: 
(i) Aggregated-level results: we first examine how robust are the cantonal specific relative risk curves to the selection of the weights, by comparing our population specific weights, with variance specific weights (i.e. for municipalities belonging to the same canton, we weight the coefficients on the basis functions by the inverse of sample variance), mimicking the meta-analyses performed in the two-stage approach \citep{GASPARRINI2022e557}.
(ii) Prior specification on seasonality: we rerun the model with a less restricted prior on day of year, by letting the probability that the standard deviation of random effect across day of year is larger than 1 equal to 0.01, while keeping all the other model specifications the same as that in the main analysis.
(iii) Comparison with frequentist approach: To compare our results with the traditional frequentist approach, which allows the temporal trend (across year and day of year) to be independent across regions, we add two unstructured residuals - $\xi_{tm}$ and $\eta_{dm}$ ($d$, $t$, and $m$ are the index for day, year and municipality respectively) - following zero-mean Gaussian process, into the main model and derive the main results. 
(iv) Comparison with the SB-DLNM approach: To compare our approach with the SB-DLNM one, we conduct a sensitivity analysis between 3 models.
The SB-DLNM, our approach and another approach combining SB-DLNM and our approach. 
A qualitative summary of the main differences between the models is provided in Supplementary Table 1.
We run the three models with the dataset in Switzerland for 2,145 municipalities from June to August, 2018-2022.
}

\section{Results}\label{result}
\subsection{Exploratory data analysis}
\begin{figure}[htbp]
    \centering
    \includegraphics[scale=0.45]{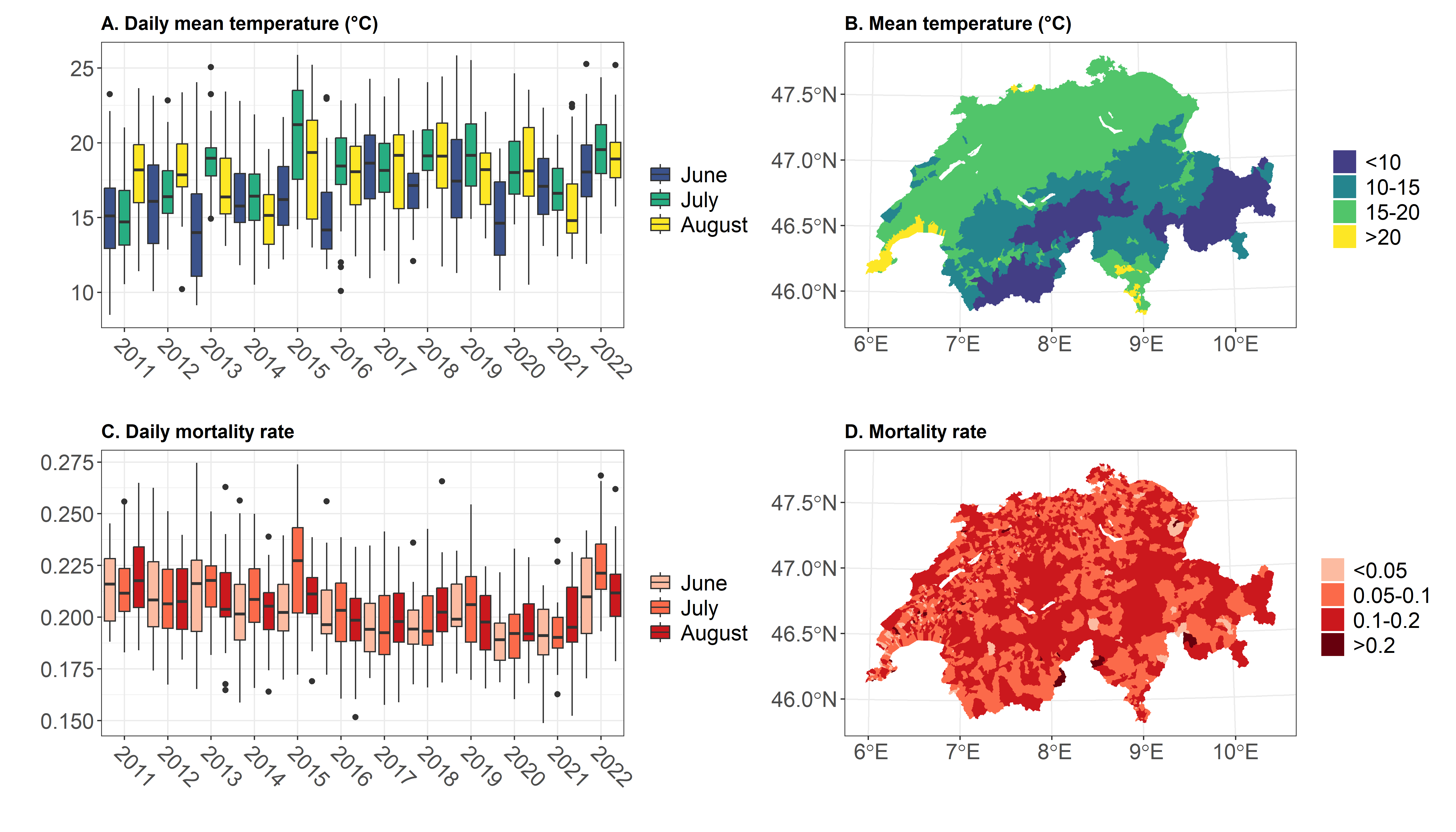}
    \caption{{\color{black}\textbf{A.}} Daily mean temperature during summers from 2011 to 2022; \textbf{B.} Spatial pattern of the mean temperature exposure across Swiss municipalities; {\color{black}\textbf{C.} }Daily mortality rate among people over 65 years old during summers from 2011 to 2022; \textbf{D.} Spatial pattern of the mortality rate among people over 65 years old across Swiss municipalities.}
    \label{Figure_1}
\end{figure}

The distributions of temperature and mortality among people older than 65 years is showed in Figure \ref{Figure_1} across the summer months during 2011-2022 in Switzerland. 
There is an increasing trend in the mean daily temperature in the study period (panel A).  
For people over 65 years old, the mortality rate (panel C) decreases over the years, with a spike after 2020, which can be explained by the COVID-19 pandemic \citep{owidcoronavirus}. 
Some seasonality patterns exist in both temperature and mortality rate time series plots.  
Looking at the spatial variability of temperature (panel B), we see that the average temperature in north-west part of Switzerland is higher than that in south, central or east areas, reflecting the mountain regions, while for mortality rate (panel D), there is not a distinct spatial pattern.

\subsection{Municipality-level relative mortality risk of temperature}
Panel A of Figure \ref{Figure_2} shows the median and 95\% CrIs of the overall (J-shape) effect of temperature on the relative mortality risk nationwide in Switzerland. 
There is weak evidence to support a temperature effect on all-cause mortality for temperatures lower than 15$^{\circ} C$. 
In contrast, there is a steep increase in the mortality risk for temperatures higher than 20$^{\circ} C$. 
Panel B of Figure \ref{Figure_2} shows the median relative risk of temperature on mortality in each municipality, reflecting the variation of temperature effects at fine spatial scale. It is evident that for high temperatures, the spatial variation of the relative risk exceeds the uncertainty bands of the nationwide one.

\subsection{Minimum mortality temperature (MMT)}
{\color{black}
Panel C of Figure \ref{Figure_2} shows the distribution of MMT in each municipality across Switzerland. 
The nationwide MMT in Switzerland is 15.24$^{\circ} C$.
%The overall mean MMT across municipalities in Switzerland is 16.55$^{\circ} C$ during summer months.
% The MMT values are relatively lower in the south part of Switzerland, corresponding to the mountainous areas with low temperature exposures.
% In comparison, the MMT values are higher in northern areas. 
Panel D of Figure \ref{Figure_2} shows the spatial distribution of MMT percentile (MMP) in each municipality. The MMP is lower in the north-west of Switzerland, whereas higher in the central and east regions, except Ticino. 
}

\subsection{Excess mortality rates attributable to heat (ERH)}
\begin{figure}[htbp]
    \centering
\includegraphics[scale = 0.5]{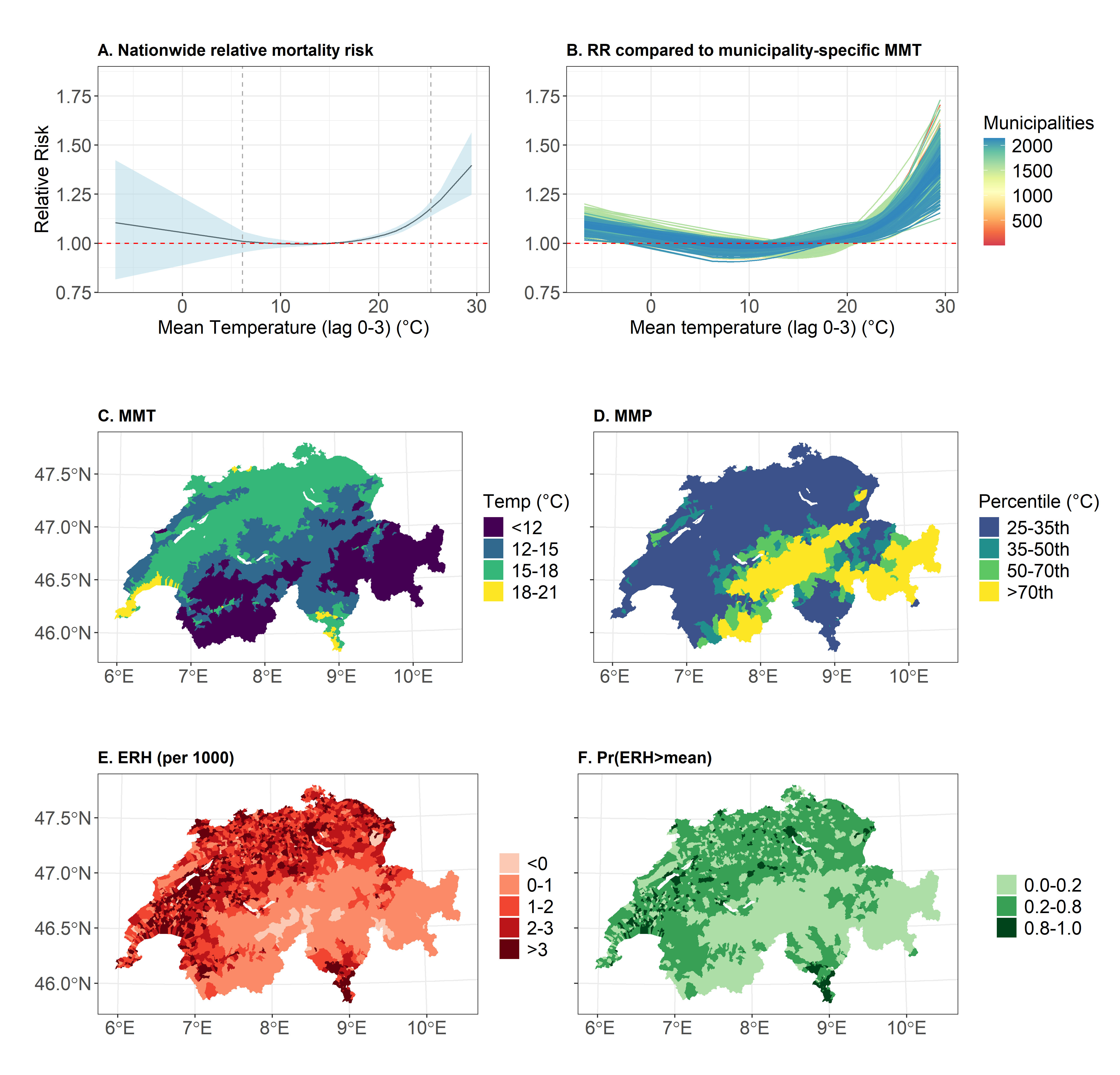}
  
    \caption{\textbf{A.} Nationwide temperature effect on mortality in Switzerland (the black curve represents the median estimate of the mortality relative risk (RR) compared to the risk at {\color{black} the nationwide minimum mortality temperature (MMT)}, the blue shaded area represents the 95\% CrI (credible interval) of the RR, {\color{black} and the dashed grey line represents the 1st and 99th percentile of the nationwide temperature distribution}); \textbf{B.} Spaghetti plot of the temperature-related median mortality risk, relative to the risk at {\color{black} the region-specific MMTs}, in each municipality;  {\color{black} \textbf{C.} MMT in each municipality; \textbf{D.} Minimum mortality temperature percentile (MMP) in each municipality; \textbf{E.} Excess mortality rates attributable to heat (ERH) per thousand population in each municipality; \textbf{F.} The exceedance probability of ERH in the area is higher than the mean value of the ERH (2.45) across the country.}} 
    \label{Figure_2}
\end{figure}

Across the country, the overall excess deaths attributable to summer heat exposure is {\color{black} 4,446 (95\% CrI:3,792 to 5,135)} across 12 years, with an average of nearly {\color{black}370} excess death counts each year in Switzerland.
Panel {\color{black}E} of Figure \ref{Figure_2} shows that the excess mortality rate is higher in the north-west of Switzerland, with values reaching more than $2$ per thousand population in {\color{black}most of the} municipalities. 
Panel {\color{black}F} of Figure \ref{Figure_2} shows the exceedance probability that ERH is greater than the mean ERH across the country, and the strongest evidence corresponds to the places characterised by the highest ERH values.

{\color{black} Supplementary Figure 2 shows the spatial distribution of the AFH. 
Across the 12 years, the overall fraction of death attributable to heat is 2.78\% (95\% CrI: 2.37\% to 3.21\%). 
In most municipalities in central Switzerland, the median AFH is less than 2\%.
While higher AFH values are observed in the northern, western and central-south parts Switzerland, where they reached 
$>10\%$ of total summer deaths in some municipalities. }
Additionally, Supplementary Figure 2 presents higher {\color{black} AFH} values (panel A) and stronger evidence (panel B) in some urban areas, such as Zurich, Basel and Geneva.

\subsection{Aggregated cantonal-level results}
The estimated cantonal-level relative mortality risk associated with temperature exposure are presented in Figure \ref{Figure_3}, while the cantonal-level J-shape relationships are shown in Supplementary Figure 3.
Panel A of Figure \ref{Figure_3} shows that {\color{black}Basel-Stadt} is the canton where the mortality risk increases most steeply when the temperature is higher than the canton-specific MMT. 
We also observe higher risk in {\color{black} Ticino, Valais, and Graubunden} mostly reflecting the French speaking part of Switzerland. 
In contrast, {\color{black} Geneva} has the lowest risk of death due to heat exposure. 
{\color{black}
Panel B of Figure \ref{Figure_3} shows the posterior probability of the relative risk being higher than the overall mean relative risk of heat (when temperature is higher than the canton-specific MMT) across all cantons, capturing the uncertainty of the estimates. 
We report strong evidence to support that the canton of Valais is the canton affected the most by heat exposure.
}

\begin{figure}[htbp]
    \centering
    \includegraphics[scale = 0.5]{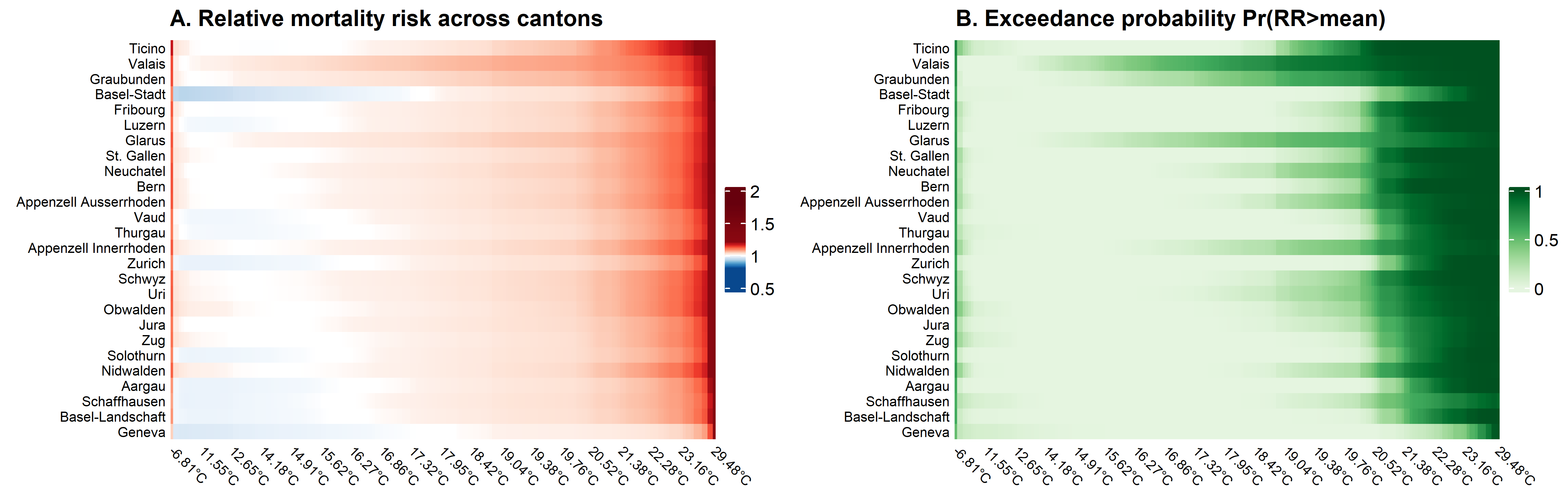}
    \caption{\textbf{A.} The median value of mortality relative risk (RR), compared to the risk at {\color{black} cantonal-specific MMTs}, in each canton; \textbf{B.} The exceedance probability of RR is higher than the mean value (1.04) across the country in each canton.}
    \label{Figure_3}
\end{figure}

\subsection{Spatial effect modifiers}
\begin{figure}[htbp]
    \centering
    \includegraphics[scale = 0.4]{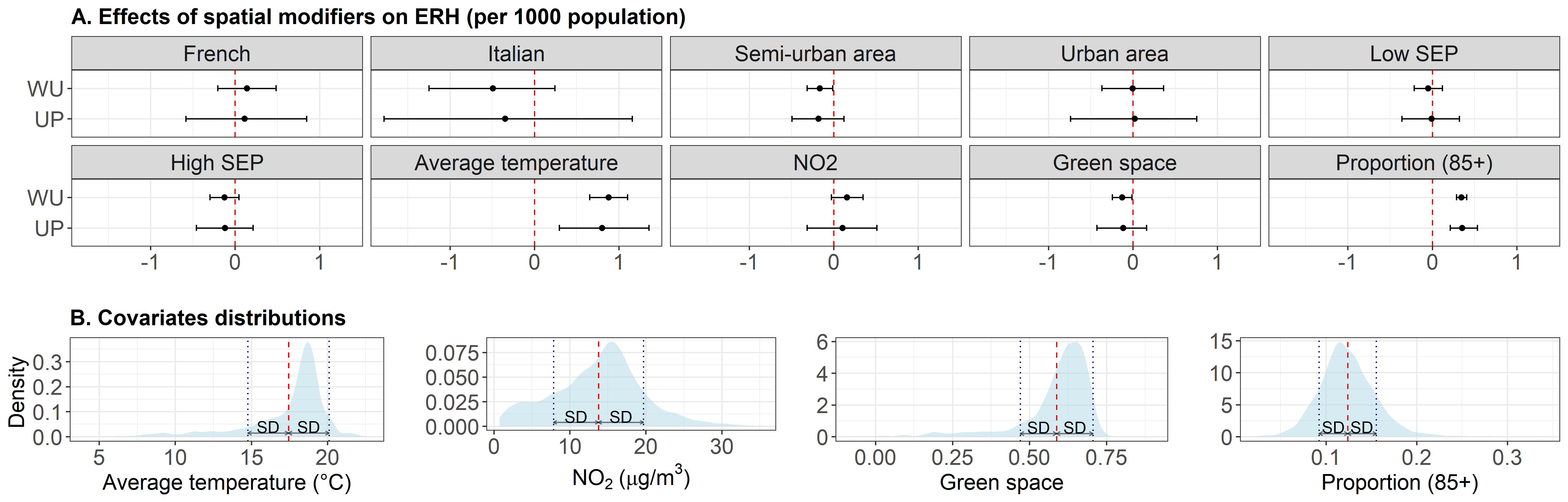}
    \caption{{\color{black}\textbf{A.}} The effects of covariates on heat-related excess mortality rate (ERH) using without uncertainty (WU) and uncertainty propagation (UP) approaches; {\color{black}\textbf{B.}} The distribution of continuous spatial effect modifiers (the shaded blue areas represent the probability density distributions of the variables, the red dashed lines show the mean of each variable, and the dark blue dotted lines show the standard deviation (SD) of each variable).}
    \label{Figure_4}
\end{figure}

Panel A of Figure \ref{Figure_4} shows the effect of spatial effect modifiers on the ERH, with and without uncertainty propagation. 
After propagating the uncertainty, {\color{black}one standard deviation increases in the proportion of people older than 85 years (3.16\%) correspond to $0.35$ (95\% CrI: $0.21$ to $0.53$)} more deaths per 1,000 people due to heat exposure. 
Access to green space was also found to affect the {\color{black}observed spatial difference in excess death rates attributable} to heat exposure. 
{\color{black}One standard deviation increases in green space coverage (0.12 of the NDVI) are associated with $0.12$ (95\% CrI: $-0.16$ to $0.43$)} less deaths per 1,000 people due to heat exposure. 
Areas with higher average temperature exposure are characterised with higher excess mortality rate, {\color{black}with one standard deviation (2.67$^\circ C$) increases in the average temperature leading to $0.80$ (95\% CrI: $0.30$ to $1.35$)} more deaths per 1,000 people due to acute heat exposure.
{\color{black} For air pollution, we observe that one standard deviation (59.15, unit $\mu g/m^3 \times 10$) of NO2 is associated with $0.10$ (95\% CrI: $-0.31$ to $0.51$) more deaths per 1,000 population attributable to heat exposure.}
The distributions of the continuous effect modifiers and their respective standard deviations are shown in panel B of Figure \ref{Figure_4}.
There is weak evidence of an association between the other covariates and excess mortality rate attributable to heat exposure.
The results of the model without uncertainty propagation are consistent, but, as expected, the CrIs are narrower (panel A of Figure \ref{Figure_4}).

\subsection{Sensitivity analysis}
{\color{black} 
When we consider variance-specific weights for aggregating the municipality specific coefficients of the basis function (sensitivity analysis i), the results are identical, Supplementary Figure 4.
Our results are robust when we use a less restrictive prior on seasonality (sensitivity analysis ii) and when we use spatiotemporal interactions to align our approach more closely with the frequentist DLNM framework (sensitivity analysis iii), see Supplementary Figures 5 and 6.
For the comparison with the SB-DLNM approach (sensitivity analysis iv), we list the differences between models in Supplementary Table 1, and find the two primary distinctions lie in the specification of the basis functions and the adjustment for spatiotemporal confounding. 
Supplementary Figure 7 shows the variation in relative risk across regions with the three models. 
Overall, we find that the SB-DLNM approach provides wigglier results with respect to the dose-response relationship and larger uncertainties. 
Finally, Supplementary Figure 8 illustrates results for three regions with varying levels of urbanisation - urban (Zurich), semi-urban (Bellinzona), and rural (Val-de-Ruz). 
In all three cases, the SB-DLNM approach yields wider 95\% CrIs than our model, reflecting greater model complexity and reduced precision.
}

\section{Discussion}\label{discussion}
This study introduces a flexible modelling framework to account for the spatial variation in the non-linear relationships between heat and mortality {\color{black}at very high spatial scale}.
By leveraging the posterior distribution of the model parameters, the approach derives key epidemiological metrics, including excess mortality attributable to heat exposure and minimum mortality temperature, to comprehensively assess the health impacts of heat. 
Set within a Bayesian framework, the approach fully propagates the uncertainty associated with model parameters into the derived epidemiological metrics and obtains aggregated estimates at any administrative level, such as cantons, deemed particularly relevant for policymakers. 
Lastly, we extend our analysis by incorporating a second-stage model to investigate the role of environmental and social effect modifiers in shaping spatial patterns of {\color{black}excess mortality rates}.

\subsection{Comparison with previous epidemiological studies}

Our results are consistent with previous epidemiological studies focusing on estimating the temperature-mortality relationships \citep{multicountryAmbient,WICKI2024118116,Masselot854}. 
We report approximately {\color{black} 370} deaths in people older than 65 years per summer in Switzerland {\color{black} and 2.78\% of deaths attributable to heat exposures}. 
{\color{black} This is compatible with the 95\% CrI with a previous study focusing on excess mortality during the summer 2022 in Switzerland \citep{konstantinoudis2023bayesian,HeatMort2022Eu,2022Switz}.}
The observed J-shape relationship is consistent with the previous studies in Japan, USA, Italy, Australia \citep{multicountryAmbient}, and Switzerland \citep{2022Switz}. 
{\color{black} The minimum mortality temperature (MMT) estimates are consistent with a global wide study on MMTs \citep{mappingMMT}, a multi-country study across European cities \citep{Masselot854}, and a nationwide study in Switzerland \citep{Swiss_nationalwide}.
Our findings of approximately 2.8\% of deaths attributable to heat exposure is in line with the studies assessing the effect of heat in European countries \citep{Masselot854} and in Switzerland \citep{2022Switz,SwissAFANRisk}.
}
However, in line with a previous study in England and Wales, we observed significant spatial variation of the MMT distribution across small areas \citep{GASPARRINI2022e557}.

Previous studies have reported spatial {\color{black} differences in heat-associated health impact} across small areas \citep{COPD_uk,bennet2014,GASPARRINI2022e557,Masselot854}.
{\color{black} A multi-country study across 854 European cities revealed the spatial disparity of excess death rates due to heat exposure across urban characteristics \citep{Masselot854}.}
Studies in England and Wales found higherheat-related mortality {\color{black}burden} in urban areas and older populations \citep{GASPARRINI2022e557,bennet2014}.
In contrast, another study in England focusing on chronic 
obstructive pulmonary disease hospital admissions reported weak evidence of a protective effect of green space and average temperature \citep{COPD_uk}. 
In line with the previous studies, we also report higher {\color{black} heat-related mortality burden among} older populations and evidence of a protective effect of green space. 
This can be explained by the impaired mechanisms of older populations to regulate heat, and also by the fact that green spaces modify heat exposure \citep{systematicReview_Vulner,greenspace}.
However, we observe that average temperature increases heat-related {\color{black}excess mortality rates}. 
Higher averaged temperature can be a proxy for higher air pollution exposure which could directly or indirectly, by aggravating chronic diseases, can increase vulnerability to heat \citep{ZHANG2024120023}.
{\color{black} 
Consistent with previous studies in California (US) \citep{AirPollution_Heat_California}, China \citep{combinedEffect_CHN,interactiveEffect_Jiangsu}, and five European countries \citep{ZHANG2024120023}, we also find positive associations between higher air pollution levels and heat-related health outcomes.
}

\subsection{Comparison with previous models}

Previous studies leveraging data from multiple countries and cities worldwide, with no explicit spatial structure that needs to be incorporated, have employed a time-series approach that accounts for lags and non-linear temperature effects \citep{DLNMgaspA,2022Switz,multicountryAmbient,Masselot854,multivarMeta_nonlinear}. 
These analyses are typically conducted independently at the first stage, with the resulting estimates pooled using meta-regression. 
This methodology has been also applied to nationwide and multi-county studies using smaller areas with prominent spatial structures \citep{GASPARRINI2022e557,2022Switz,HeatMort2022Eu,Masselot854}. 
When the main unit of analysis is small areas, the above approach encounters two interconnected challenges: data sparsity at the small-area level and the lack of a component that captures spatial dependencies. 
Sparse data can lead to unstable estimates, characterised by high variability, while not accounting for spatial correlation brings variance underestimation \citep{challgeospat}. 
Using spatial priors that incorporate spatial structures can smooth estimates, providing more stable and reliable results in the presence of sparse data \citep{ReducingmetaDLNM,childhoodcancer_Swiss}.

Previous studies have also developed linear threshold models to capture the J-shaped relationship between heat exposure and mortality \citep{COPD_uk, bennet2014}. These models incorporate spatial priors to account for spatial dependencies in the slope above the threshold \citep{COPD_uk, bennet2014}. While the results are easy to communicate and interpret, these models face challenges in identifying thresholds, and two linear segments are might not fully capture the non-linear effect of temperature on health.

A recent study in Barcelona proposed a spatial Bayesian distributed lag non-linear model {\color{black}(SB-DLNM)} framework that accounts for the spatial dependencies of the non-linear temperature-mortality relationship in 73 neighborhoods in Barcelona \citep{sb-dlnm}. 
{\color{black} 
When comparing municipality-level relative risks, the SB-DLNM approach produced more variable dose-response curves with larger uncertainties. 
This reflects its use of higher-order interactions and highly flexible basis functions. 
In our context, such complexity is unnecessary as seasonal mortality patterns in Switzerland show limited variation by year or small region. 
Higher-order interactions can also overadjust, which is a possible explanation of the decreasing trend of the relative risk in higher temperatures. 
The basis function of SB-DLNM is designed to capture the non-linear lagged effects of temperature. 
As we are focusing on quantifying the impacts of heat, averaging temperatures over lag 0-3 days is adequate \citep{COPD_uk,bennet2014,modLagSimu_gas}.}

This model provides a first step in disentangling the non-linear spatial inequalities of heat exposure on health. 
In our approach, we extend this model in several ways. 
{\color{black} 
First, we scaled the framework to fit large epidemiological datasets and examined variable aspects of the mortality-heat relationship. 
Focusing on Switzerland, we used 12 years' worth of data, 2,145 small areas and the INLA algorithm for quick and accurate inference \citep{inlaref}. 
We exploited the Bayesian framework to show different epidemiological metrics, and different policy-relevant aggregations, while fully propagating parameter uncertainty. 
Second, we also proposed a second stage model to quantify the potential sources of the observed spatial inequalities by selected small-area environmental and social covariates. }
{\color{black}Third}, we introduced a population offset and a spatial random effect in the model. 
The population offset is crucial in small-area studies as population can confound the exposure response relationship. 
In case of heat exposure, small areas with higher populations, especially within large cities, can be proxies for factors that confound the temperature-mortality relationship such as buildings, green space access and air pollution. 
{\color{black}Fourth}, we included a spatial random effect to account for remaining unobserved spatial factors that confound the heat-mortality relationship, such as deprivation. 
{\color{black}Additionally}, we selected different priors. 
For the spatial random effect, we selected a BYM2 prior instead of the Leroux prior. 
The marginal variance of the spatial field in the Leroux prior depends on the graph structure. 
The BYM2 model is defined by standardising the spatial and non-spatial random effects, facilitating interpretable hyperprior assignment and hyperparameter interpretation (ensures the same interpretation in different applications) \citep{BYMref,Leroux2000EstimationOD,SORBYE201439}. 
{\color{black}Finally}, we also used PC priors to ensure the parsimony of the model and avoid the risk of overfitting according to the distance from the simplest base model \citep{Gómezbook}.

Our work has some limitations. 
First, although we explained the spatial vulnerability with environmental and socio-demographic variables, the spatial variation of heat-related vulnerability can also be explained by other individual-level variables, such as medication and chronic conditions. 
Such granular information was not available. 
Second, while the analysis incorporated data from multiple years, the model accounts only for the spatially varying effects of temperature. 
The assumption of a constant heat-mortality relationship over time might not hold, as previous studies have reported adaptive mechanisms to heat over time \citep{konstantinoudis2023asthma}. 
Within the proposed framework, it is straightforward to let the coefficients of the basis function to vary in time, using priors with temporal structures, such as random walks and autoregressive processes \citep{VICEDOCABRERA2018239}. 
However, we chose not to pursue this approach, as the primary objective of this study was to disentangle spatial inequalities. 
Disregarding temporal variation is unlikely to substantially impact the results while offering a computationally less intensive model. 
{\color{black}Third}, rather than including temperature exposures at individual lags to capture delayed effects, we calculated the average temperature across 0-3 time lags, as shorter lags are relevant when assessing heat exposure. 
However, when examining exposures with longer lags, it would be important to extend this approach to account for potential non-linear structures in the lag dimension.
{\color{black} Finally, it is important to note that the spatial effect modifiers do not necessarily represent spatial vulnerability to heat exposure per se. 
Rather, as they are influenced by baseline mortality rates, they may partially capture underlying geographic variation in all-cause mortality unrelated to heat-specific susceptibility. }

In summary, in this work, we present a scalable framework to describe the spatial variation of the heat-mortality effect across small areas. 
The framework can be expanded to capture spatial, temporal and spatiotemporal vulnerabilities of the health-related burden of temperature exposure. 
Applied to Switzerland, our framework provides valuable insights for mitigation strategies and the reduction of spatial inequalities in the health impacts of heat exposure. 
We emphasize the significance of older populations and green space in shaping spatial vulnerabilities, as well as the marked spatial variation in the MMTs. 
Consequently, we suggest targeted interventions for older populations, region-specific heat warnings that incorporate varying MMTs, and a greater emphasis on green space in urban planning.

\section*{Code and data availability} \sloppy
R code is publicly available on GitHub {\color{black} \url{https://github.com/fxinyichen/SwissHeat_svc}.
Data delivery is governed by a contract with the Swiss Federal Statistical Office (FSO). For further details, please contact the FSO directly: \url{https://www.bfs.admin.ch/bfs/en/home/services/contact.html}. To support reproducibility, we provide a simulated dataset to demonstrate the analysis in \url{https://doi.org/10.5281/zenodo.16923676}.}

\section*{Competing interests}
The authors declare no competing interests.

\section*{Author contributions statement}
G.K. conceived the study. G.K. and M.B. supervised the study. G.K. prepared the mortality, population and covariate data. X.C. developed the statistical model, wrote the code for the analysis and extracted the results. X.C. wrote the initial draft and all the authors contributed in modifying the paper and critically interpreting the results. All authors read and approved the final version for publication.

\section*{Acknowledgments}
G.K. is supported by an Imperial College Research Fellowship. 

%%%%%%%%%%%%%%

\bibliographystyle{apalike}
\bibliography{references.bib}{}

\begin{thebibliography}{}

\bibitem[Arsad et~al., 2022]{systematicReview_Vulner}
Arsad, F., Hod, R., Ahmad, N., Ismail, R., Mohamed, N., Baharom, M., Osman, Y., Radi, M., and Tangang, F. (2022).
\newblock The impact of heatwaves on mortality and morbidity and the associated vulnerability factors: A systematic review.
\newblock {\em \textit{International Journal of Environmental Research and Public Health}}, 19:16356.
\newblock \url{https://doi.org/10.3390/ijerph192316356}.

\bibitem[Baccini et~al., 2008]{heatLag03}
Baccini, M., Biggeri, A., Accetta, G., Kosatsky, T., Katsouyanni, K., Analitis, A., Anderson, H., Bisanti, L., D'Ippoliti, D., Danova, J., Forsberg, B., Medina, S., Paldy, A., Rabczenko, D., Schindler, C., and Michelozzi, P. (2008).
\newblock Heat effects on mortality in 15 european cities.
\newblock {\em \textit{Epidemiology (Cambridge, Mass.)}}, 19:711--719.
\newblock \url{https://doi.org/10.1097/EDE.0b013e318176bfcd}.

\bibitem[Ballester et~al., 2023]{HeatMort2022Eu}
Ballester, J., Quijal-Zamorano, M., Turrubiates, R., Pegenaute, F., Herrmann, F., Robine, J.-M., Basagaña, X., Tonne, C., Antó, J., and Achebak, H. (2023).
\newblock Heat-related mortality in europe during the summer of 2022.
\newblock {\em \textit{Nature Medicine}}, 29:1--10.
\newblock \url{https://doi.org/10.1038/s41591-023-02419-z}.

\bibitem[Bennett et~al., 2014]{bennet2014}
Bennett, J.~E., Blangiardo, M., Fecht, D., Elliott, P., and Ezzati, M. (2014).
\newblock {Vulnerability to the mortality effects of warm temperature in the districts of England and Wales}.
\newblock {\em \textit{Nature Climate Change}}, 4(4):269--273.
\newblock \url{https://ideas.repec.org/a/nat/natcli/v4y2014i4d10.1038_nclimate2123.html}.

\bibitem[Besag et~al., 1991]{BYMref}
Besag, J., York, J., and Mollié, A. (1991).
\newblock Bayesian image restoration, with two applications in spatial statistics.
\newblock {\em \textit{Annals of the Institute of Statistical Mathematics}}, 71:1--59.
\newblock \url{https://doi.org/10.1007/BF00116466}.

\bibitem[Dai et~al., 2024]{combinedEffect_CHN}
Dai, W., Liu, S., Xu, W., Shen, Y., Yang, X., and Zhou, Q. (2024).
\newblock The combined effects of heatwaves, air pollution and greenery on the risk of frailty: a national cohort study.
\newblock {\em Scientific reports}, 14(1):24293.

\bibitem[de~Schrijver et~al., 2021]{tempGrid}
de~Schrijver, E., Folly, C.~L., Schneider, R., Royé, D., Franco, O.~H., Gasparrini, A., and Vicedo-Cabrera, A.~M. (2021).
\newblock A comparative analysis of the temperature-mortality risks using different weather datasets across heterogeneous regions.
\newblock {\em GeoHealth}, 5(5):e2020GH000363.

\bibitem[de~Schrijver et~al., 2023]{Swiss_nationalwide}
de~Schrijver, E., Sivaraj, S., Raible, C., Franco, O., Chen, K., and Vicedo-Cabrera, A. (2023).
\newblock Nationwide projections of heat- and cold-related mortality impacts under climate change and population development scenarios in switzerland.
\newblock {\em \textit{Environmental Research Letters}}.
\newblock \url{https://iopscience.iop.org/article/10.1088/1748-9326/ace7e1}.

\bibitem[Ebi et~al., 2021]{EBI2021698}
Ebi, K.~L., Capon, A., Berry, P., Broderick, C., {de Dear}, R., Havenith, G., Honda, Y., Kovats, R.~S., Ma, W., Malik, A., Morris, N.~B., Nybo, L., Seneviratne, S.~I., Vanos, J., and Jay, O. (2021).
\newblock Hot weather and heat extremes: health risks.
\newblock {\em \textit{The Lancet}}, 398(10301):698--708.
\newblock \url{https://www.sciencedirect.com/science/article/pii/S0140673621012083}.

\bibitem[{eurostat}, 2011]{Swiss_urban}
{eurostat} (2011).
\newblock \textit{Degree of urbanisation classification}.
\newblock \url{https://ec.europa.eu/eurostat/statistics-explained/index.php?title=Degree_of_urbanisation_classification_-_2011_revision} [Accessed 23rd October 2024].

\bibitem[{Federal Statistical Office}, 2010]{Swiss_lang}
{Federal Statistical Office} (2010).
\newblock \textit{Spatial divisions}.
\newblock \url{https://www.bfs.admin.ch/bfs/de/home/statistiken/querschnittsthemen/raeumliche-analysen/raeumliche-gliederungen.html} [Accessed 23rd October 2024].

\bibitem[{Federal Statistical Office}, 2024]{SwissPop}
{Federal Statistical Office} (2024).
\newblock \textit{Population}.
\newblock \url{https://www.bfs.admin.ch/bfs/en/home/statistics/population.html} [Accessed 14th February 2024].

\bibitem[Gasparrini, 2011]{DLNM_R}
Gasparrini, A. (2011).
\newblock Distributed lag linear and non-linear models in {R}: The package \texttt{dlnm}.
\newblock {\em Journal of Statistical Software}, 43(8):1--20.

\bibitem[Gasparrini, 2016]{modLagSimu_gas}
Gasparrini, A. (2016).
\newblock Modelling lagged associations in environmental time series data: A simulation study.
\newblock {\em Epidemiology (Cambridge, Mass.)}, 27(6):835—842.

\bibitem[Gasparrini, 2022]{ts_ttlGas}
Gasparrini, A. (2022).
\newblock A tutorial on the case time series design for small-area analysis.
\newblock {\em \textit{BMC Medical Research Methodology}}, 22.
\newblock \url{https://api.semanticscholar.org/CorpusID:248498944}.

\bibitem[Gasparrini and Armstrong, 2010]{tsDesign}
Gasparrini, A. and Armstrong, B. (2010).
\newblock Time series analysis on the health effects of temperature: Advancements and limitations.
\newblock {\em \textit{Environmental research}}, 110:633--8.
\newblock \url{https://doi.org/10.1016/j.envres.2010.06.005}.

\bibitem[Gasparrini et~al., 2010]{DLNMgaspA}
Gasparrini, A., Armstrong, B., and Kenward, M.~G. (2010).
\newblock Distributed lag non-linear models.
\newblock {\em \textit{Statistics in Medicine}}, 29(21):2224--2234.
\newblock \url{https://onlinelibrary.wiley.com/doi/abs/10.1002/sim.3940}.

\bibitem[Gasparrini et~al., 2012]{multivarMeta_nonlinear}
Gasparrini, A., Armstrong, B., and Kenward, M.~G. (2012).
\newblock Multivariate meta-analysis for non-linear and other multi-parameter associations.
\newblock {\em \textit{Statistics in Medicine}}, 31(29):3821--3839.
\newblock \url{https://onlinelibrary.wiley.com/doi/abs/10.1002/sim.5471}.

\bibitem[Gasparrini and Ben, 2013]{ReducingmetaDLNM}
Gasparrini, A. and Ben, A. (2013).
\newblock Reducing and meta-analysing estimates from distributed lag non-linear models.
\newblock {\em BMC Medical Research Methodology}, 13(1).

\bibitem[Gasparrini et~al., 2015]{multicountryAmbient}
Gasparrini, A., Guo, Y., Hashizume, M., Lavigne, E., Zanobetti, A., and Schwartz, J. (2015).
\newblock Mortality risk attributable to high and low ambient temperature: a multicountry observational study.
\newblock {\em \textit{Lancet}}, 386.
\newblock \url{https://doi.org/10.1016/S0140-6736(14)62114-0}.

\bibitem[Gasparrini et~al., 2022]{GASPARRINI2022e557}
Gasparrini, A., Masselot, P., Scortichini, M., Schneider, R., Mistry, M.~N., Sera, F., Macintyre, H.~L., Phalkey, R., and Vicedo-Cabrera, A.~M. (2022).
\newblock Small-area assessment of temperature-related mortality risks in england and wales: a case time series analysis.
\newblock {\em \textit{The Lancet Planetary Health}}, 6(7):e557--e564.
\newblock \url{https://www.sciencedirect.com/science/article/pii/S2542519622001383}.

\bibitem[Gómez~Rubio, 2020]{Gómezbook}
Gómez~Rubio, V. (2020).
\newblock {\em \textit{Bayesian Inference with INLA}}.
\newblock Chapman and Hall/CRC.
\newblock \url{https://doi.org/10.1201/9781315175584}.

\bibitem[Koldasbayeva et~al., 2024]{challgeospat}
Koldasbayeva, D., Tregubova, P., Gasanov, M., Zaytsev, A., Petrovskaia, A., and Burnaev, E. (2024).
\newblock Challenges in data-driven geospatial modeling for environmental research and practice.
\newblock {\em Nature Communications}, 15.

\bibitem[Konstantinoudis et~al., 2021]{5EUCOVID2020excess}
Konstantinoudis, G., Cameletti, M., Gómez~Rubio, V., León, I., Pirani, M., Baio, G., Larrauri, A., Riou, J., Egger, M., Vineis, P., and Blangiardo, M. (2021).
\newblock Regional excess mortality during the 2020 covid-19 pandemic: a study of five european countries.

\bibitem[Konstantinoudis et~al., 2023a]{konstantinoudis2023bayesian}
Konstantinoudis, G., Hauser, A., and Riou, J. (2023a).
\newblock Bayesian ensemble modelling to monitor excess deaths during summer 2022 in switzerland.
\newblock {\em arXiv preprint arXiv:2308.15251}.

\bibitem[Konstantinoudis et~al., 2023b]{konstantinoudis2023asthma}
Konstantinoudis, G., Minelli, C., Lam, H. C.~Y., Fuertes, E., Ballester, J., Davies, B., Vicedo-Cabrera, A.~M., Gasparrini, A., and Blangiardo, M. (2023b).
\newblock Asthma hospitalisations and heat exposure in england: a case--crossover study during 2002--2019.
\newblock {\em Thorax}, 78(9):875--881.

\bibitem[Konstantinoudis et~al., 2022]{COPD_uk}
Konstantinoudis, G., Minelli, C., Vicedo-Cabrera, A.~M., Ballester, J., Gasparrini, A., and Blangiardo, M. (2022).
\newblock Ambient heat exposure and copd hospitalisations in england: a nationwide case-crossover study during 2007-2018.
\newblock {\em \textit{Thorax}}, 77(11):1098—1104.
\newblock \url{https://europepmc.org/articles/PMC9606528}.

\bibitem[Konstantinoudis et~al., 2019]{childhoodcancer_Swiss}
Konstantinoudis, G., Schuhmacher, D., Ammann, R., Diesch-Furlanetto, T., Kuehni, C., and Spycher, B. (2019).
\newblock Bayesian spatial modelling of childhood cancer incidence in switzerland using exact point data: A nationwide study during 1985-2015.
\newblock {\em International Journal of Health Geographics}.

\bibitem[Leroux et~al., 2000]{Leroux2000EstimationOD}
Leroux, B., Lei, X., and Breslow, N. (2000).
\newblock Estimation of disease rates in small areas: A new mixed model for spatial dependence.
\newblock {\em Institute for Mathematics and Its Applications}, 116.

\bibitem[Madaniyazi et~al., 2023]{shouldAdjust}
Madaniyazi, L., Tobías, A., Vicedo-Cabrera, A.~M., Jaakkola, J. J.~K., Honda, Y., Guo, Y., Schwartz, J., Zanobetti, A., Bell, M.~L., Armstrong, B., Campbell, M.~J., Katsouyanni, K., Haines, A., Ebi, K.~L., Gasparrini, A., and Hashizume, M. (2023).
\newblock Should we adjust for season in time-series studies of the short-term association between temperature and mortality?
\newblock {\em \textit{Epidemiology (Cambridge, Mass.)}}, 34(3):313—318.
\newblock \url{https://doi.org/10.1097/ede.0000000000001592}.

\bibitem[Markevych et~al., 2017]{greenspace}
Markevych, I., Schoierer, J., Hartig, T., Chudnovsky, A., Hystad, P., Dzhambov, A., de~Vries, S., Triguero-Mas, M., Brauer, M., Nieuwenhuijsen, M., Lupp, G., Richardson, E., Astell-Burt, T., Dimitrova, D., Feng, X., Sadeh, M., Standl, M., and Fuertes, E. (2017).
\newblock Exploring pathways linking greenspace to health: Theoretical and methodological guidance.
\newblock {\em \textit{Environmental Research}}, 158:301--317.
\newblock \url{https://doi.org/10.1016/j.envres.2017.06.028}.

\bibitem[Masselot et~al., 2023]{Masselot854}
Masselot, P., Mistry, M., Vanoli, J., Schneider, R., Iungman, T., Garcia-Leon, D., Ciscar, J.-C., Feyen, L., Orru, H., Urban, A., Breitner, S., Huber, V., Schneider, A., Samoli, E., Stafoggia, M., de'Donato, F., Rao, S., Armstrong, B., Nieuwenhuijsen, M., Vicedo-Cabrera, A.~M., Gasparrini, A., {MCC Collaborative Research Network}, and {EXHAUSTION project} (2023).
\newblock Excess mortality attributed to heat and cold: a health impact assessment study in 854 cities in europe.
\newblock {\em \textit{The Lancet. Planetary health}}, 7(4):e271--e281.
\newblock \url{http://repositori.upf.edu/bitstream/10230/57139/1/Masselot_lph_exce.pdf}.

\bibitem[Mathieu et~al., 2020]{owidcoronavirus}
Mathieu, E., Ritchie, H., Rodés-Guirao, L., Appel, C., Giattino, C., Hasell, J., Macdonald, B., Dattani, S., Beltekian, D., Ortiz-Ospina, E., and Roser, M. (2020).
\newblock Coronavirus pandemic (covid-19).
\newblock {\em \textit{Our World in Data}}.
\newblock \url{https://ourworldindata.org/coronavirus}.

\bibitem[Meng et~al., 2021]{Meng2021NO2_MortalityCardRespir}
Meng, X., Liu, C., Chen, R., Sera, F., Vicedo-Cabrera, A.~M., Milojevic, A., Guo, Y., Tong, S., Coelho, M. S. Z.~S., Saldiva, P. H.~N., Lavigne, E., Correa, P.~M., Ortega, N.~V., Osorio, S., Garcia, ., and Kan, H. (2021).
\newblock Short term associations of ambient nitrogen dioxide with daily total, cardiovascular, and respiratory mortality: multilocation analysis in 398 cities.
\newblock {\em BMJ (Clinical research ed.)}, 372:n534.

\bibitem[{MeteoSwiss}, 2020]{SwissTempData}
{MeteoSwiss} (2020).
\newblock \textit{Spatial Climate analysis}.
\newblock \url{https://www.meteoswiss.admin.ch/climate/the-climate-of-switzerland/spatial-climate-analyses.html} [Accessed 14th February 2024].

\bibitem[{Meteotest}, 2010]{Swiss_NO2}
{Meteotest} (2010).
\newblock \textit{Air Quality}.
\newblock \url{https://meteotest.ch/en/division/luftreinhaltung} [Accessed 3rd June 2025].

\bibitem[Moraga, 2019]{PCpriorBook}
Moraga, P. (2019).
\newblock {\em \textit{Geospatial Health Data: Modeling and Visualization with R-INLA and Shiny}}.
\newblock Chapman \& Hall/CRC Biostatistics Series.
\newblock \url{https://www.paulamoraga.com/book-geospatial/index.html}.

\bibitem[{Nager.Date}, 2024]{NagerDate}
{Nager.Date} (2024).
\newblock \textit{Worldwide public holiday}.
\newblock \url{https://date.nager.at/} [Accessed 19th June 2024].

\bibitem[Panczak et~al., 2012]{Swiss_socioeconomic}
Panczak, R., Galobardes, B., Voorpostel, M., Spoerri, A., Zwahlen, M., and Egger, M. (2012).
\newblock A swiss neighbourhood index of socioeconomic position: Development and association with mortality.
\newblock {\em \textit{Journal of epidemiology and community health}}, 66.
\newblock \url{https://doi.org/10.1136/jech-2011-200699}.

\bibitem[Quijal-Zamorano et~al., 2024]{sb-dlnm}
Quijal-Zamorano, M., Martinez-Beneito, M.~A., Ballester, J., and Marí-Dell’Olmo, M. (2024).
\newblock {Spatial Bayesian distributed lag non-linear models (SB-DLNM) for small-area exposure-lag-response epidemiological modelling}.
\newblock {\em \textit{International Journal of Epidemiology}}, 53(3):dyae061.
\newblock \url{https://doi.org/10.1093/ije/dyae061}.

\bibitem[{R Core Team}, 2024]{R}
{R Core Team} (2024).
\newblock {\em R: A Language and Environment for Statistical Computing}.
\newblock R Foundation for Statistical Computing, Vienna, Austria.
\newblock \url{https://www.R-project.org/}.

\bibitem[Riebler et~al., 2016]{BYMvsLeroux}
Riebler, A., Sørbye, S., Simpson, D., and Rue, H. (2016).
\newblock An intuitive bayesian spatial model for disease mapping that accounts for scaling.
\newblock {\em Statistical Methods in Medical Research}, 25:1145--1165.

\bibitem[Rue et~al., 2009]{inlaref}
Rue, H., Martino, S., and Chopin, N. (2009).
\newblock Approximate bayesian inference for latent gaussian models by using integrated nested laplace approximations.
\newblock {\em \textit{Journal of the Royal Statistical Society Series B}}, 71:319--392.
\newblock \url{https://doi.org/10.1111/j.1467-9868.2008.00700.x}.

\bibitem[Saki et~al., 2020]{NO2_copd}
Saki, H., Goudarzi, G., Jalali, S., Barzegar, G., Farhadi, M., Parseh, I., Geravandi, S., Salmanzadeh, S., Yousefi, F., and Mohammadi, M.~J. (2020).
\newblock Study of relationship between nitrogen dioxide and chronic obstructive pulmonary disease in bushehr, iran.
\newblock {\em Clinical Epidemiology and Global Health}, 8(2):446--449.

\bibitem[Schulte et~al., 2024]{SwissAFANRisk}
Schulte, F., Röösli, M., and Ragettli, M.~S. (2024).
\newblock Risk, attributable fraction and attributable number of cause-specific heat-related emergency hospital admissions in switzerland.
\newblock {\em International Journal of Public Health}, Volume 69 - 2024.

\bibitem[Simpson et~al., 2017]{PriorSimpson}
Simpson, D., Rue, H., Riebler, A., Martins, T.~G., and S{\o}rbye, S.~H. (2017).
\newblock {Penalising Model Component Complexity: A Principled, Practical Approach to Constructing Priors}.
\newblock {\em Statistical Science}, 32(1):1 -- 28.

\bibitem[Sun et~al., 2020]{AirPollution_Heat_California}
Sun, Y., Ilango, S., Schwarz, L., Wang, Q., Chen, J.-C., Lawrence, J., Wu, J., and Benmarhnia, T. (2020).
\newblock Examining the joint effects of heatwaves, air pollution, and green space on the risk of preterm birth in california.
\newblock {\em Environmental Research Letters}, 15:104099.

\bibitem[{Swiss Data Cube}, 2020]{Swiss_NDVI}
{Swiss Data Cube} (2020).
\newblock \textit{Normalized Difference Vegetation Index (NDVI) - Annual Mean - Switzerland}.
\newblock \url{https://yareta.unige.ch/archives/b6022b1c-1c59-4fc7-8a76-1b68c3c07dc7} [Accessed 23rd October 2024].

\bibitem[Sørbye and Rue, 2014]{SORBYE201439}
Sørbye, S.~H. and Rue, H. (2014).
\newblock Scaling intrinsic gaussian markov random field priors in spatial modelling.
\newblock {\em Spatial Statistics}, 8:39--51.
\newblock Spatial Statistics Miami.

\bibitem[Tobias et~al., 2021]{MMTglobal}
Tobias, A., Hashizume, M., Honda, Y., Sera, F., Ng, C. F.~S., Kim, Y., Royé, D., Chung, Y., Tran, N.~D., Kim, H., Lee, W., Íñiguez, C., Vicedo-Cabrera, A., Abrutsky, R., Guo, Y., Tong, S., Coelho, M., Saldiva, P., Lavigne, E., and Carrasco-Escobar, G. (2021).
\newblock Geographical variations of the minimum mortality temperature at a global scale.
\newblock {\em \textit{Environmental Epidemiology}}, 5:e169.
\newblock \url{https://doi.org/10.1097/ee9.0000000000000169}.

\bibitem[Vicedo-Cabrera et~al., 2023]{2022Switz}
Vicedo-Cabrera, A., de~Schrijver, E., Schumacher, D., Ragettli, M., Fischer, E., and Seneviratne, S. (2023).
\newblock The footprint of human-induced climate change on heat-related deaths in the summer of 2022 in switzerland.
\newblock {\em \textit{Environmental Research Letters}}, 18.
\newblock \url{https://doi.org/10.1088/1748-9326/ace0d0}.

\bibitem[Vicedo-Cabrera et~al., 2018]{VICEDOCABRERA2018239}
Vicedo-Cabrera, A.~M., Sera, F., Guo, Y., Chung, Y., Arbuthnott, K., Tong, S., Tobias, A., Lavigne, E., {de Sousa Zanotti Stagliorio Coelho}, M., {Hilario Nascimento Saldiva}, P., Goodman, P.~G., Zeka, A., Hashizume, M., Honda, Y., Kim, H., Ragettli, M.~S., Röösli, M., Zanobetti, A., Schwartz, J., Armstrong, B., and Gasparrini, A. (2018).
\newblock A multi-country analysis on potential adaptive mechanisms to cold and heat in a changing climate.
\newblock {\em Environment International}, 111:239--246.

\bibitem[Wicki et~al., 2024]{WICKI2024118116}
Wicki, B., Flückiger, B., Vienneau, D., {de Hoogh}, K., Röösli, M., and Ragettli, M.~S. (2024).
\newblock Socio-environmental modifiers of heat-related mortality in eight swiss cities: A case time series analysis.
\newblock {\em \textit{Environmental Research}}, 246:118116.
\newblock \url{https://www.sciencedirect.com/science/article/pii/S0013935124000203}.

\bibitem[Yang et~al., 2025]{Yang2025NOxMortality}
Yang, S., Li, M., Guo, C., and et~al. (2025).
\newblock Associations of long-term exposure to nitrogen oxides with all-cause and cause-specific mortality.
\newblock {\em Nature Communications}, 16:1730.

\bibitem[Yin et~al., 2019]{mappingMMT}
Yin, Q., Wang, J., Ren, Z., Li, J., and Guo, Y. (2019).
\newblock Mapping the increased minimum mortality temperatures in the context of global climate change.
\newblock {\em Nature communications}, 10(1):4640.

\bibitem[Zhang et~al., 2024]{ZHANG2024120023}
Zhang, S., Breitner, S., Stafoggia, M., de' Donato, F., Samoli, E., Zafeiratou, S., Katsouyanni, K., Rao, S., {Diz-Lois Palomares}, A., Gasparrini, A., Masselot, P., Nikolaou, N., Aunan, K., Peters, A., and Schneider, A. (2024).
\newblock Effect modification of air pollution on the association between heat and mortality in five european countries.
\newblock {\em Environmental Research}, 263:120023.

\bibitem[Zhou et~al., 2022]{interactiveEffect_Jiangsu}
Zhou, L., Wang, Y., Wang, Q., Ding, Z., Jin, H., Zhang, T., and Zhu, B. (2022).
\newblock The interactive effects of extreme temperatures and pm2.5 pollution on mortalities in jiangsu province, china.

\bibitem[Èrica Martínez-Solanas et~al., 2021]{MARTINEZSOLANAS2021e446}
Èrica Martínez-Solanas, Quijal-Zamorano, M., Achebak, H., Petrova, D., Robine, J.-M., Herrmann, F.~R., Rodó, X., and Ballester, J. (2021).
\newblock Projections of temperature-attributable mortality in europe: a time series analysis of 147 contiguous regions in 16 countries.
\newblock {\em \textit{The Lancet Planetary Health}}, 5(7):e446--e454.
\newblock \url{https://www.sciencedirect.com/science/article/pii/S2542519621001509}.

\end{thebibliography}

\end{document}

\maketitle
\newpage
% \tableofcontents
%\listoftables
\renewcommand{\listfigurename}{List of Supplementary Figures}
\renewcommand{\listtablename}{List of Supplementary Table}
\listoffigures
\newpage
\listoftables
\newpage
\renewcommand{\figurename}{Supplementary Figure} 
\renewcommand{\tablename}{Supplementary Table} 
    
\begin{figure}[htbp]
\caption{Map of the cantons in Switzerland.}
    \centering
\includegraphics[scale = 0.7]{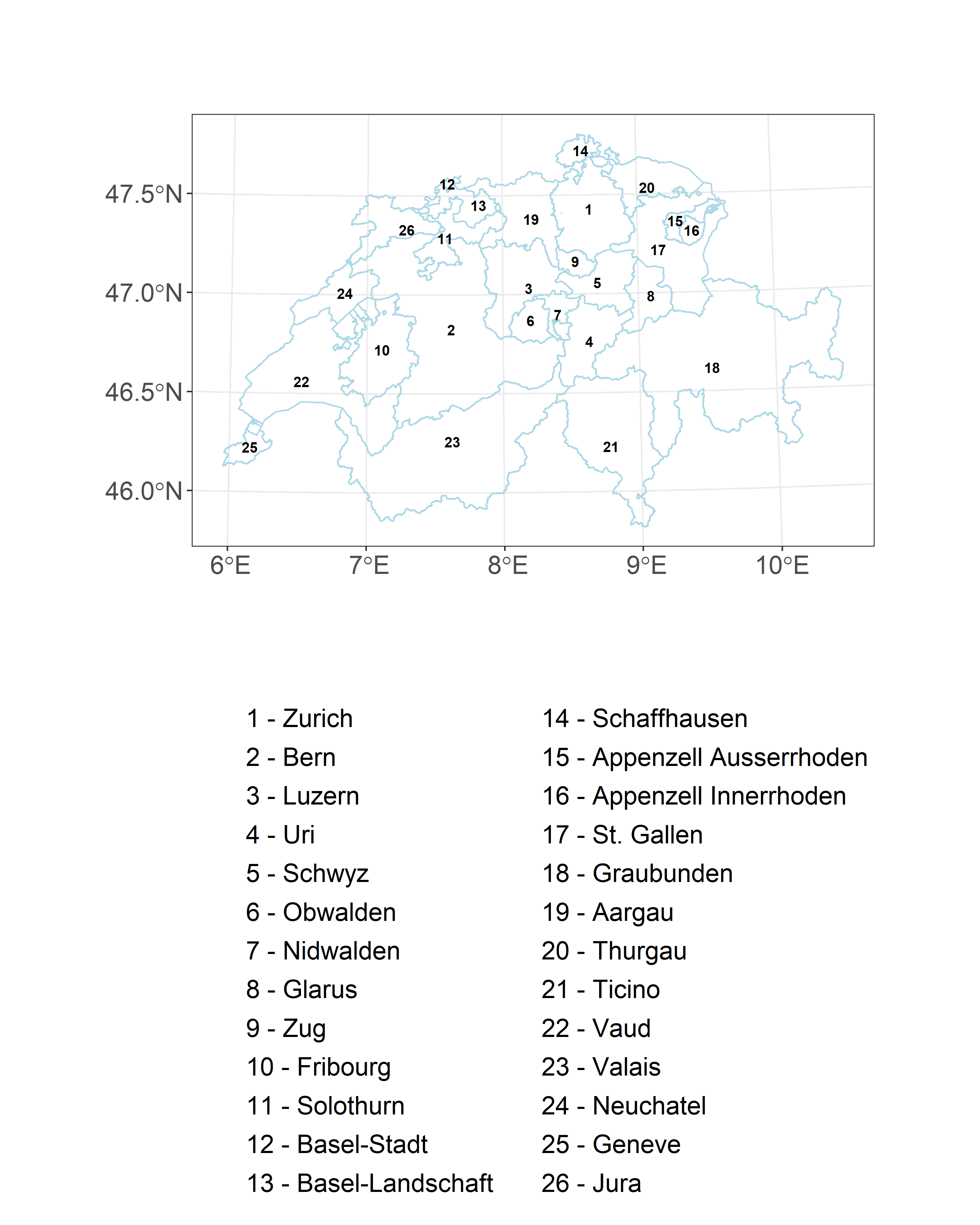}

    \label{SuppleFigure_1}
\end{figure}

\newpage

\begin{figure}[htbp]
    \caption{{\color{black}\textbf{A.}} Fraction of deaths attributable to heat {\color{black}(AFH)} (when temperature is higher than the {\color{black}region-specific MMT}) in each municipality; {\color{black}\textbf{B.}} The exceedance probability of {\color{black}AFH} in the area is higher than the mean value across the country.}
    \centering
    \includegraphics[scale = 0.55]{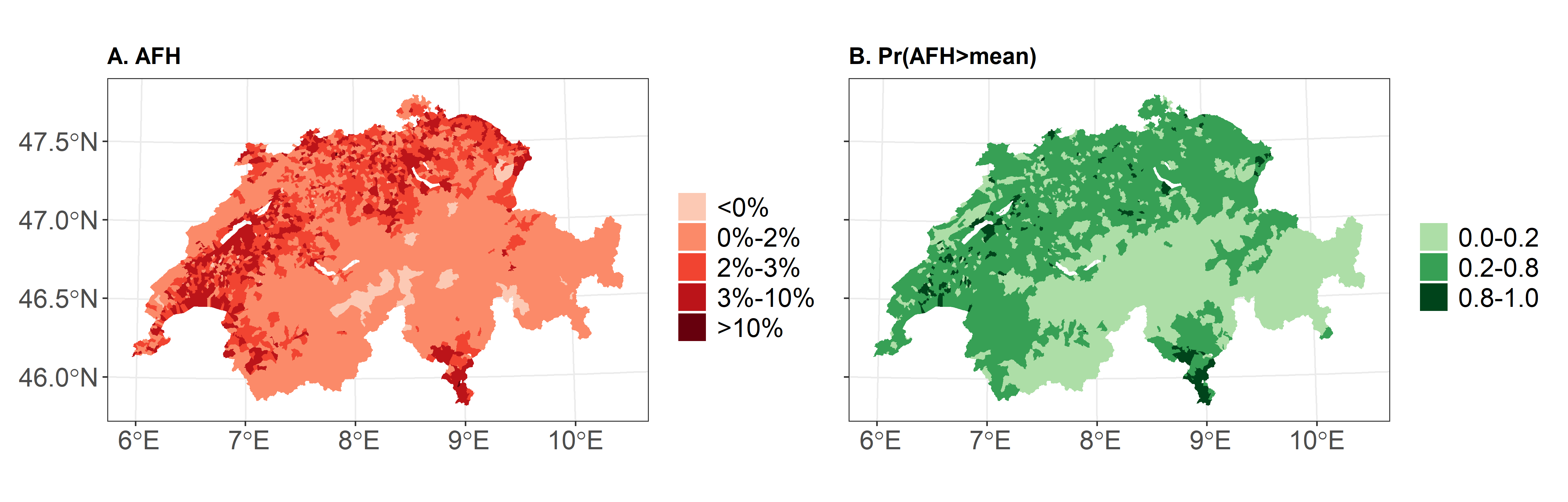}
    \label{SuppleFigure_2}
\end{figure}

\newpage

\begin{figure}[htbp]
    \caption{{\color{black}Relative mortality risk in each canton, compared to the canton-specific MMT, in Switzerland.}}
    \centering
    \includegraphics[scale = 0.4]{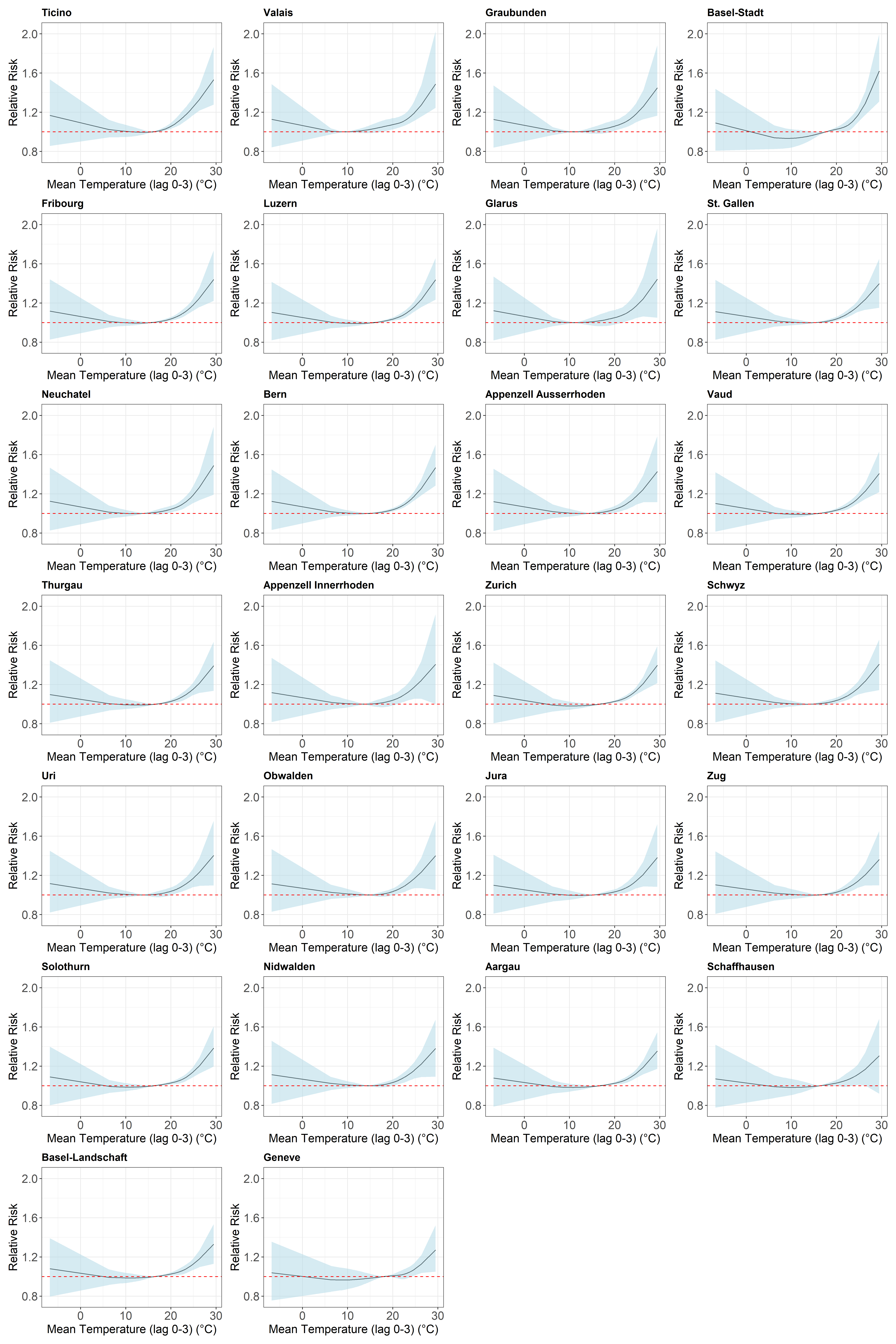}

    \label{SuppleFigure_3}
\end{figure}

\newpage
\begin{figure}[htbp]
    \caption{{\color{black}
    \textbf{A.} The median value of mortality relative risk (RR), using the population-weighted approach, compared to the risk at canton-specific MMT, in each canton; \textbf{B.} The exceedance probability of population-weighted RR is higher than the mean value (1.04)  of relative risk when temperature is higher than the region-specific MMT across the country in each canton; \textbf{C.} The median value of RR, using the variance-weighted approach, compared to the risk at canton-specific MMT, in each canton; \textbf{D.} The exceedance probability of variance-weighted RR is higher than the mean value (1.04)  of relative risk when temperature is higher than the region-specific MMT across the country in each canton.}
    }
    \centering
    \includegraphics[scale = 0.55]{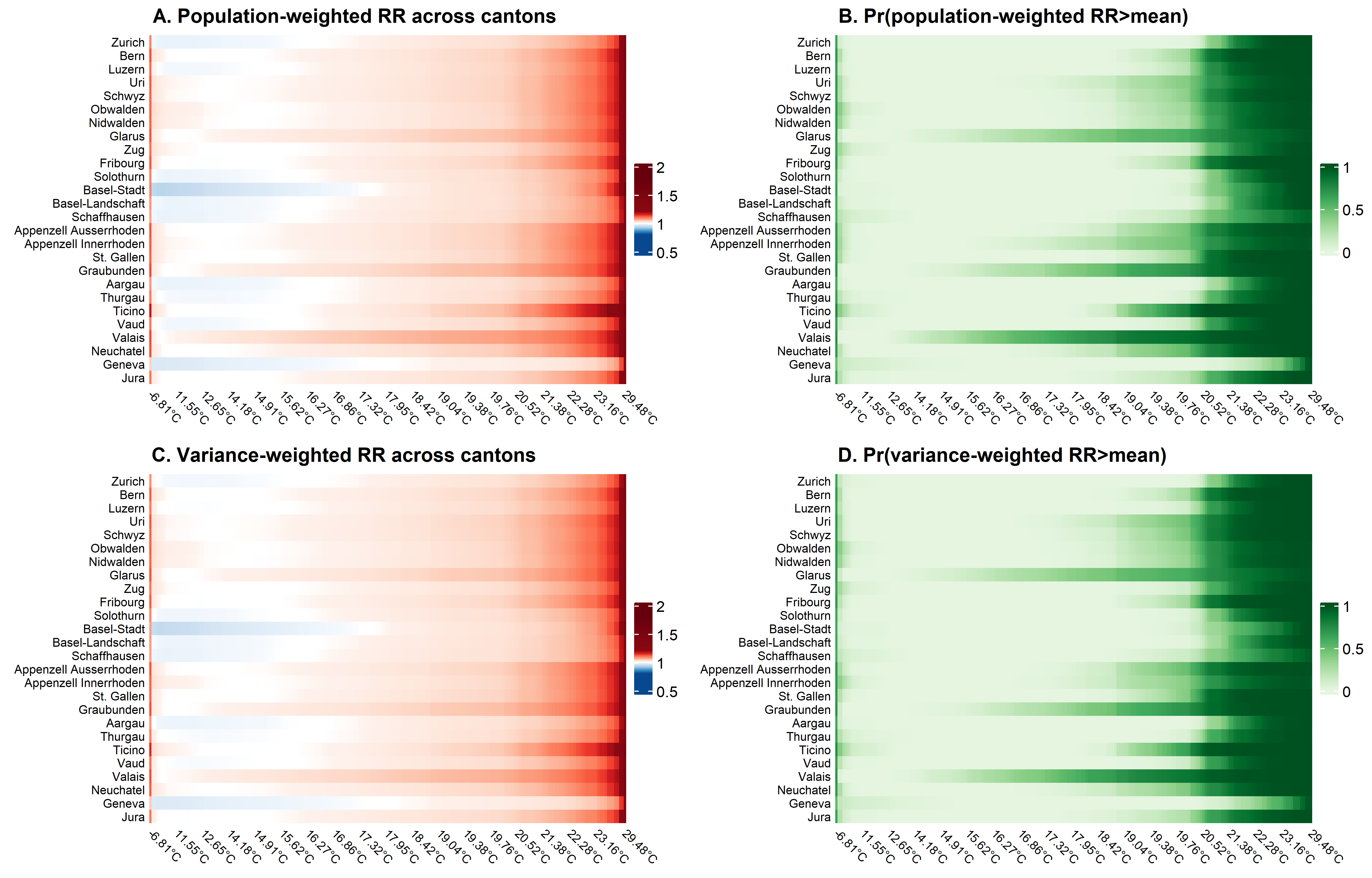}
    \label{SuppleFigure_4}
\end{figure}

\newpage
\begin{figure}[ht]
    \centering
    \caption{Sensitivity analysis results with less restricted prior on day of year for \textbf{A.} Nationwide temperature effect on mortality in Switzerland (the black curve represents the median estimate of the mortality relative risk (RR) compared to the risk at {\color{black} the nationwide minimum mortality temperature (MMT) 15.24$^\circ C$}, and the blue shaded area represents the 95\% CrI (credible interval) of the RR); \textbf{B.} Spaghetti plot of the temperature-related median mortality risk, {\color{black}relative to the risk at MMT in each municipality}; {\color{black}\textbf{C.} Minimum mortality temperature (MMT) in each municipality; \textbf{D.} Minimum mortality temperature percentile (MMP) in each municipality;} \textbf{E.} Excess mortality rates attributable to heat (ERH) per thousand population in each municipality; \textbf{F.} The exceedance probability of ERH in the area is higher than the mean value of the ERH {\color{black}(2.47)} across the country.}
\includegraphics[scale = 0.5]{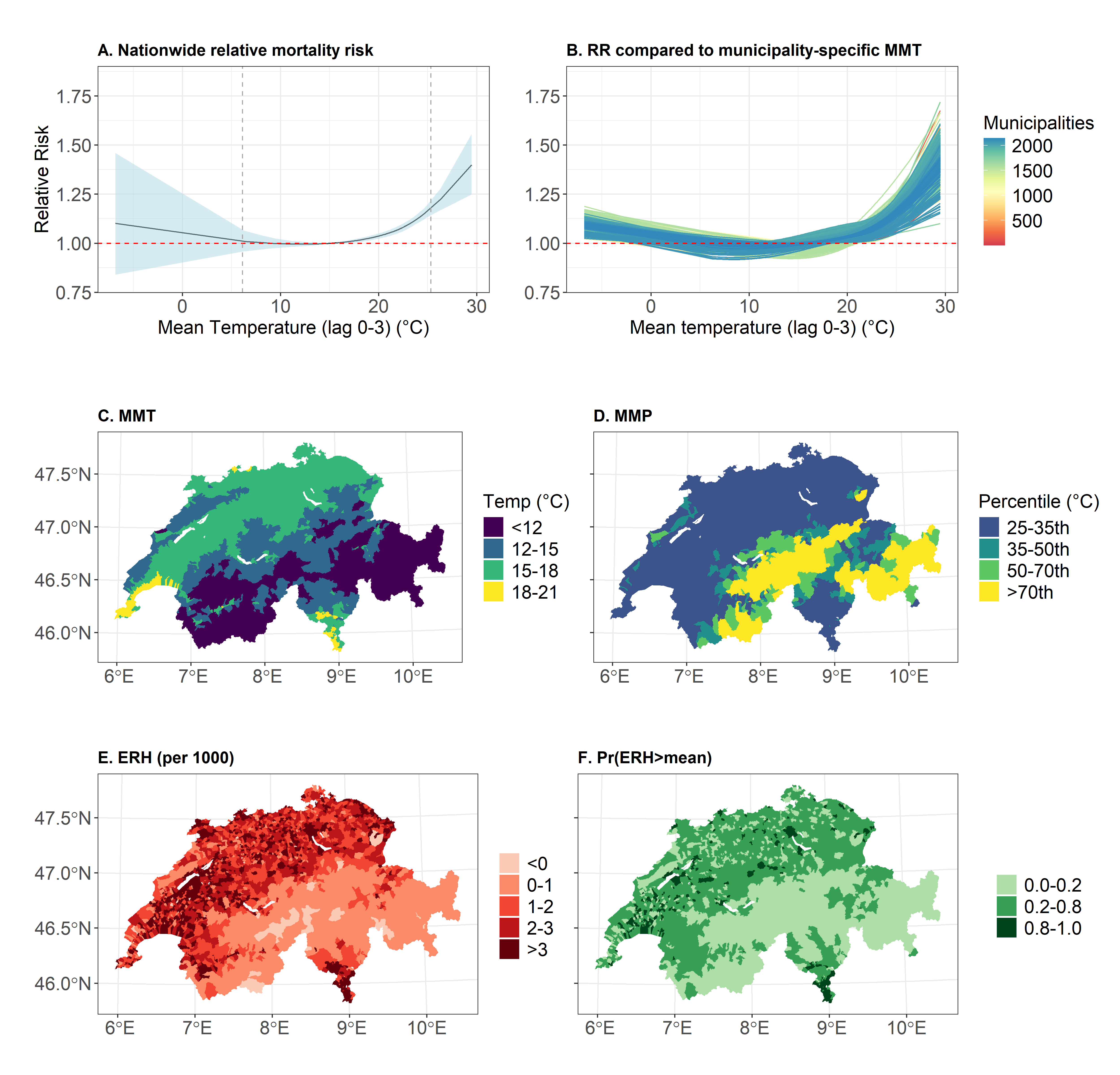}

    \label{SuppleFigure_5}
\end{figure}

\newpage
\begin{figure}[ht]
    \centering
    \caption{{\color{black}
    Sensitivity analysis results with residuals varying across time and space for \textbf{A.} Nationwide temperature effect on mortality in Switzerland (the black curve represents the median estimate of the mortality relative risk (RR) compared to the risk at nationwide minimum mortality temperature (MMT)  15.24$^\circ C$, and the blue shaded area represents the 95\% CrI (credible interval) of the RR); \textbf{B.} Spaghetti plot of the temperature-related median mortality risk, relative to the risk at MMT in each municipality; \textbf{C.} Minimum mortality temperature (MMT) in each municipality; \textbf{D.} Minimum mortality temperature percentile (MMP) in each municipality; \textbf{E.} Excess mortality rates attributable to heat (ERH) per thousand population in each municipality; \textbf{F.} The exceedance probability of ERH in the area is higher than the mean value of the ERH (2.47) across the country.}}
\includegraphics[scale = 0.5]{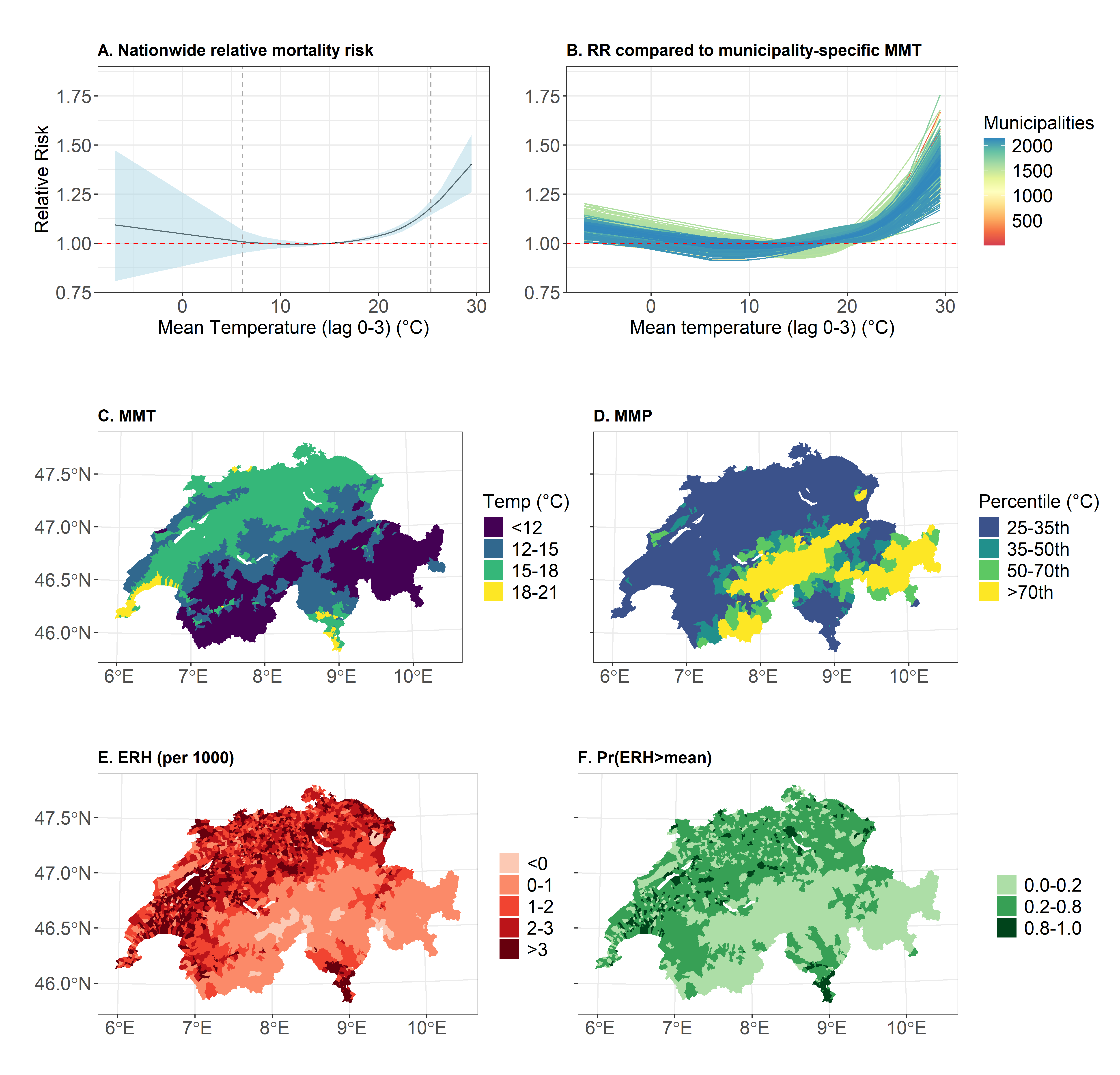}
    \label{SuppleFigure_6}
\end{figure}

\newpage

\begin{table}[]
    \caption{{\color{black}Comparison of model specification between the SB-DLNM time series approach, the approach in this paper and the combined approach}}
    \centering
    {\renewcommand{\baselinestretch}{0.6}
    \begin{tabular}{p{2cm}|p{5cm}|p{5cm}|p{4cm}}
      \hline\hline
        &SB-DLNM time-series approach & Our approach & Combined approach \\
        \hline\hline
        % Time period & 2007-2016, June to September each year&& 2011-2022, June to August each year \\
        % \hline
        % Region & 73 neighbourhoods of the city of Barcelona && 2145 municipalities in Switzerland \\
        % \hline
        Basis function & crossbasis function (using natural cubic splines with knots at the 50 and 90th of the temperature distribution; and 2 knots positioned at equally-spaced log-values using a lag period of 8 days); different basis for each region &  natural cubic splines with knots at the 10, 50 and 90th of the distribution for the mean temperature across lags (0-3); nationwide basis shared across regions& crossbasis the same as the SB-DLNM time series approach \\
        \hline
        Priors & Vague priors & PC priors & PC priors\\
        \hline
        Offset & no offset & population offset & population offset \\
         \hline
        Spatially varying effect & Leroux model&BYM2 model & BYM2 model \\
        \hline
        Spatial residual & no  &spatial residual following BYM2& spatial residual following BYM2 \\
        \hline
        DOW & intercept term varing across dow and regions &factor term & factor term \\
        \hline
        Holiday & not considered &categorical national holiday& categorical national holiday \\
        \hline
        Seasonality & day of year (doy) splines with 4 degree of freedom (df), varying across region and year& doy following RW2 in main analysis; and adding a residual term varying across region and doy following iid in sensitivity analysis&doy following RW2  \\
        \hline
        Long-term trend & date with 1 df per 10 years following iid across regions &yearly residual following iid in main analysis; and adding a residual term varying across year and region following iid in sensitivity analysis& yearly residual following iid  \\
      \hline
    \end{tabular}
    }

    \label{SuppleTable_1}
\end{table}

\begin{figure}[ht]
    \caption{{\color{black}
    \textbf{A.} The relative risk of mortality in each municipality with the SB-DLNM approach, using data from 2018-2022 in Switzerland;
    \textbf{B.}  The relative risk of mortality in each municipality with our approach, using data from 2018-2022 in Switzerland;
     \textbf{C.}  The relative risk of mortality in each municipality with the combined approach, using data from 2018-2022 in Switzerland.}}
    \centering
    \includegraphics[scale = 0.5]{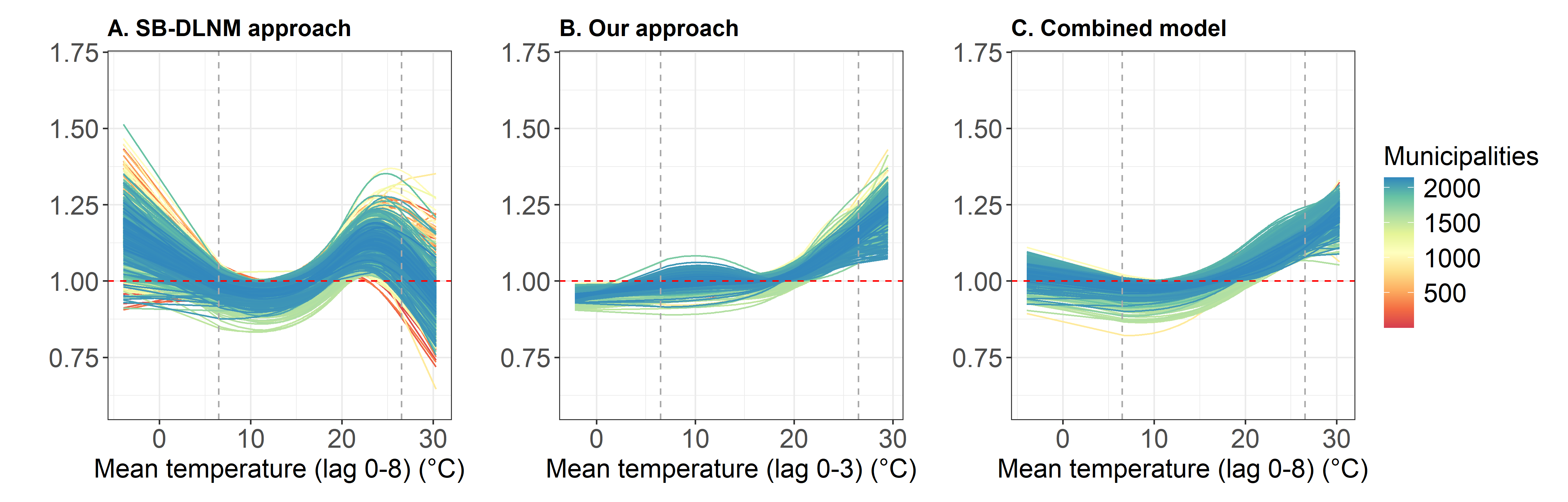}
    \label{SuppleFigure_7}
\end{figure}

\begin{figure}[ht]
    \caption{{\color{black}
     The relative risk of mortality in Zurich (\textbf{A1}), Bellinzona (\textbf{A2}), and  Val-de-Ruz (\textbf{A3}) with the SB-DLNM approach, using data from 2018-2022 in Switzerland;
     The relative risk of mortality in Zurich (\textbf{B1}), Bellinzona (\textbf{B2}), and  Val-de-Ruz (\textbf{B3}) with our approach, using data from 2018-2022 in Switzerland;
      The relative risk of mortality in Zurich (\textbf{C1}), Bellinzona (\textbf{C2}), and  Val-de-Ruz (\textbf{C3}) with the combined model, using data from 2018-2022 in Switzerland.}}

    \centering
    \includegraphics[scale = 0.5]{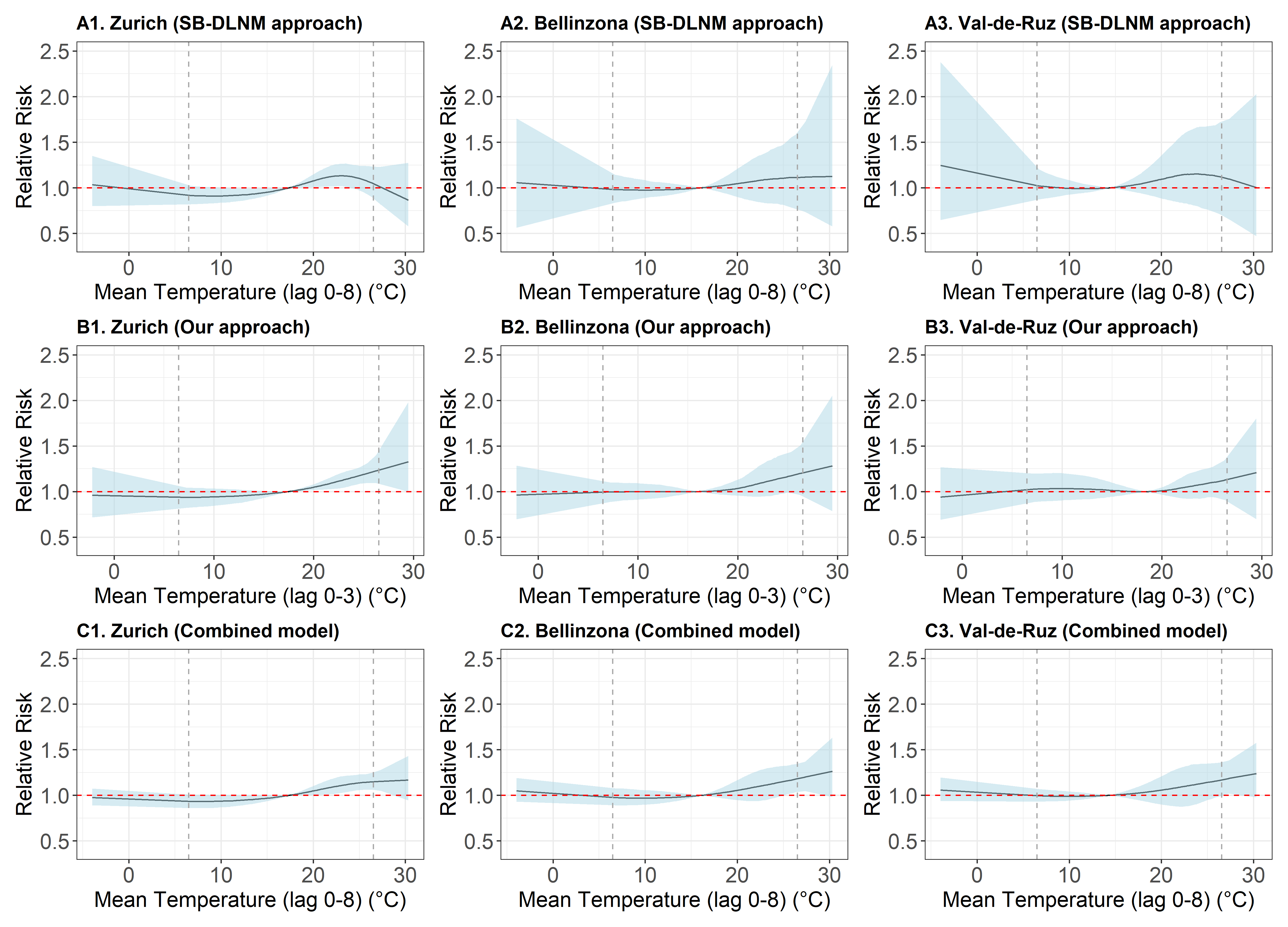}
    \label{SuppleFigure_8}
\end{figure}